\documentclass[iop,revtex]{emulateapj}

\usepackage{hyperref}
\usepackage{breakurl}
\hypersetup{
  colorlinks=true,
  urlcolor=blue,
  linkcolor=blue,
  citecolor=blue
}
\usepackage[all]{hypcap}
\usepackage{lscape}

\shorttitle{O- and B-Type Stars in W3}
\shortauthors{Kiminki et al.}
\slugcomment{{\sc Accepted to \apj:} 30 August 2015 } 

\begin{document}

\title{The O- and B-Type Stellar Population in W3: Beyond the
  High-Density Layer}

\author{Megan M. Kiminki\altaffilmark{1}, Jinyoung Serena
  Kim\altaffilmark{1}, Micaela B. Bagley\altaffilmark{1,2}, William
  H. Sherry\altaffilmark{3,4}, and George H. Rieke\altaffilmark{1}}

\altaffiltext{1}{Steward Observatory, University of Arizona, 933 North
  Cherry Avenue, Tucson, AZ 85721, USA;
  \href{mailto:mbagley@email.arizona.edu}{mbagley@email.arizona.edu}}

\altaffiltext{2}{Minnesota Institute for Astrophysics, School of
  Physics and Astronomy, University of Minnesota, 116 Church Street
  S.E., Minneapolis, MN 55455, USA}

\altaffiltext{3}{National Optical Astronomy Observatories, 950 North
  Cherry Avenue, Tucson, AZ 87719, USA}

\altaffiltext{4}{Eureka Scientific, 2452 Delmer Street Suite 100,
  Oakland, CA 94602, USA}

\begin{abstract}
  We present the first results from our survey of the star-forming
  complex W3, combining $VRI$ photometry with multiobject spectroscopy
  to identify and characterize the high-mass stellar population across
  the region.  With 79 new spectral classifications, we bring the
  total number of spectroscopically-confirmed O- and B-type stars in
  W3 to 105.  We find that the high-mass slope of the mass function in
  W3 is consistent with a Salpeter IMF, and that the extinction toward
  the region is best characterized by an $R_V$ of approximately $3.6$.
  B-type stars are found to be more widely dispersed across the W3
  giant molecular cloud (GMC) than previously realized: they are not
  confined to the high-density layer (HDL) created by the expansion of
  the neighboring W4 \ion{H}{2} region into the GMC.  This broader
  B-type population suggests that star formation in W3 began
  spontaneously up to 8--10 Myr ago, although at a lower level than
  the more recent star formation episodes in the HDL.  In addition, we
  describe a method of optimizing sky subtraction for fiber spectra in
  regions of strong and spatially-variable nebular emission.
\end{abstract}


\section{INTRODUCTION}

Massive stars have a dramatic impact on their environment, affecting
the surrounding matter distribution and energy budget through their
winds, ionizing radiation, and eventually supernovae.  They influence
processes on scales ranging from the reionization of the universe to
the timescale of circumstellar disk dispersal.  One notable effect of
massive star feedback may be the triggering of new generations of star
formation, either by sweeping neighboring molecular gas into a dense
shell which subsequently fragments into pre-stellar cores
\citep[e.g.,][]{elm77,whi94,elm98} or by compressing existing dense
clumps past the point of gravitational instability
\citep[e.g.,][]{san82,ber89}.

The W3 star-forming region \citep{wes58} is a group of \ion{H}{2}
regions and luminous infrared sources located in the outer Galaxy.
Its distance of 2.0 kpc has been well established by maser parallax
measurements \citep{hac06,xu06}.  W3 is associated with a giant
molecular cloud (GMC) of mass $4\times10^5$ M$_{\odot}$ \citep{pol12}.
The GMC borders on the extensive \ion{H}{2} region W4, which is
currently being ionized by the young cluster IC 1805 \citep{mas95}.
Considered a classic case of probable triggered star formation
\citep{lad78,car00,oey05,ruc07,fei08}, the W3/W4 complex has long been
the focus of massive-star and star-formation research \citep[see][and
  references therein]{meg08}.

\begin{figure*}
\plotone{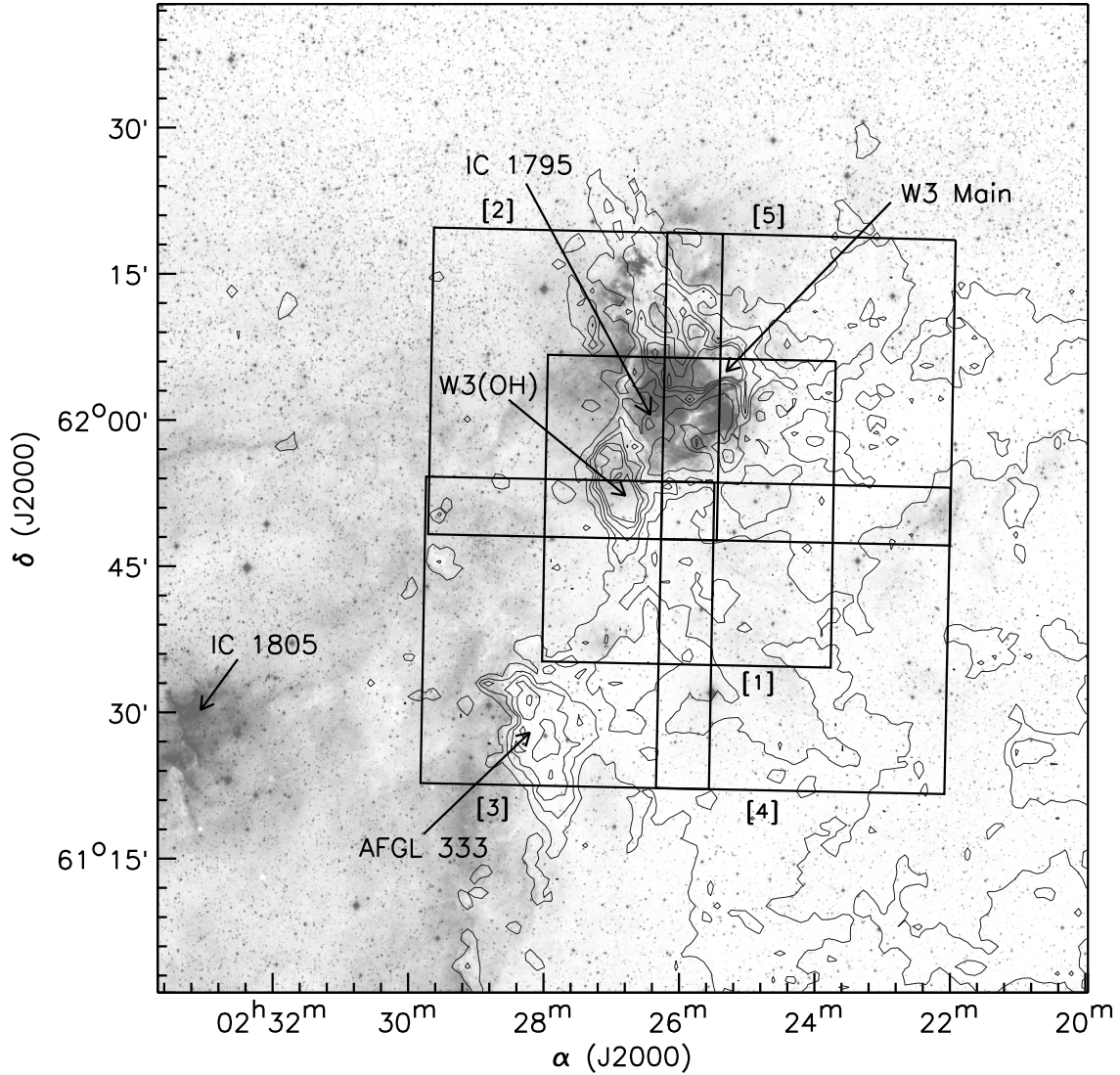}
\caption{Image of the W3/W4 complex from the Digitized Sky Survey (DSS)
  with $^{12}$CO $J= 2$--1 contours from \citet{bie11}, along
  with outlines of the five dither positions used in our 2005 90Prime
  imaging.  IC 1805 (W4) and the major features of the W3 HDL are
  labeled.
\label{fig:pointing}}
\end{figure*}

The edge of the W3 GMC has been compressed by the expanding W4
\ion{H}{2} region into a high-density layer \citep[HDL;][]{lad78},
which contains about half of the total mass of the cloud
\citep{pol12}.  Along this interface lies a sequence of star-forming
sub-regions, of which the most visibly prominent is IC 1795 (see
Figure \ref{fig:pointing}).  At 3--5 Myr old, IC 1795 is the oldest of
the clusters in W3's HDL \citep{oey05,roc11}.  Its \ion{H}{2} region
appears to have contributed to triggering the formation of its younger
neighbors, W3(OH) and W3 Main \citep{oey05}, although it has been
argued \citep{fei08} that W3 Main's spherical shape favors a
spontaneous star-formation scenario instead.  South of the other
sub-regions lies AFGL 333, another site of active star formation that
appears to have been triggered by the expansion of W4 into the W3 GMC
\citep{riv11}.

Much of the study of W3 to date has focused on the clusters in the
HDL, but recent results have begun to probe a more widespread and
relatively older stellar population.  \citet{riv11} detected groups of
relatively older young stars to the west of W3 Main and AFGL 333,
suggesting a prior generation of star formation not triggered by
feedback from W4.  Similarly, \citet{rom11} saw evidence in the
$K$-band luminosity function for an 8-Myr-old population near W3(OH).
The simple picture of the W3/W4 complex as a sequence of triggered
clusters may not be as clean as it initially appears.

In this paper, we present the first results of our survey of the full
stellar content of W3, focusing here on the high-mass population.  We
describe our photometric and spectroscopic observations (including our
approach to removing sky emission from stellar spectra) in Section
\ref{sec:obs} and discuss the identification, classification, and spatial
distribution of O- and B-type stars in Section \ref{sec:sample}.  In
Section \ref{sec:extinct}, we put constraints on the extinction law toward
W3, which we use in Section \ref{sec:HRDsect} to estimate stellar
parameters and construct an H-R diagram.  The high-mass initial mass
function (IMF) for W3 is presented in Section \ref{sec:IMFsect}.  Finally,
we discuss the implications of our results in the context of W3's
star-formation history in Section \ref{sec:disc}.

\section{OBSERVATIONS AND DATA REDUCTION}
\label{sec:obs}

\subsection{Optical Photometry}

We observed W3 with the 90Prime wide-field imager \citep{wil04} at the
Steward Observatory $2.3$-m Bok Telescope on 2005 October 7.  90Prime
consists of four 4k $\times$ 4k CCDs, each with a field of view of
$\sim 30 \arcmin \times 30 \arcmin$ and a pixel scale of $0.45
\arcsec$~pix$^{-1}$.  Due to weather constraints and CCD failure, we
were only able to use data from the single chip (Chip 2) for which we
obtained adequate standard star coverage.  Combining five dithered
observations, shown in Figure \ref{fig:pointing}, gave us roughly one
square degree of coverage in Cousins/Bessell $V$ and $I$, with dithers
1--4 also observed in $R$.  Each dither consisted of a $30$-s exposure
in $V$, a $5$-s exposure in $R$, and a $5$-s exposure in $I$, except
for dither 1 for which the $I$-band exposure was $7$ s long.  The
seeing of the observations ranged from 1.2\arcsec~to 1.9\arcsec~with a
median of 1.5\arcsec.

A more complete discussion of our imaging observations and photometry
will be the subject of an upcoming paper (J. Jose et al., in
preparation).  Briefly, we performed standard data reduction
procedures, measured photometry, and applied aperture corrections
using the IDL pipelines \emph{Bokproc}, \emph{Bokphot}, and
\emph{aperture\_correct}, which are designed for use on 90Prime images
(W. H. Sherry et al., in preparation).  Aperture corrections were fit
across each image using IDL's built-in routine for minimum curvature
surface interpolation.  Because the aperture corrections varied
strongly across 90Prime's large field of view, we processed and
measured photometry on each dithered image separately and merged the
photometric catalogs post-calibration.  The most prominant saturated
stars were automatically removed during the reduction pipeline, and
additional saturated stars and stars that overlapped diffraction
spikes or bad columns were identified manually and flagged in the
resulting catalog.  We calibrated our instrumental magnitudes using
same-night observations of the SA 92 and SA 98 standard fields
\citep{ste00}.\footnote{See also
  \url{http://www.cadc-ccda.hia-iha.nrc-cnrc.gc.ca/en/community/STETSON/standards/}.}
The standard fields were observed under comparable seeing conditions
at a range of airmasses bracketing the target airmass.  We then fit
the slope of the color term for the photometric calibration as a
function of airmass.

\begin{figure}
\includegraphics[width=84mm,trim = 8mm 0 4mm 0 ]{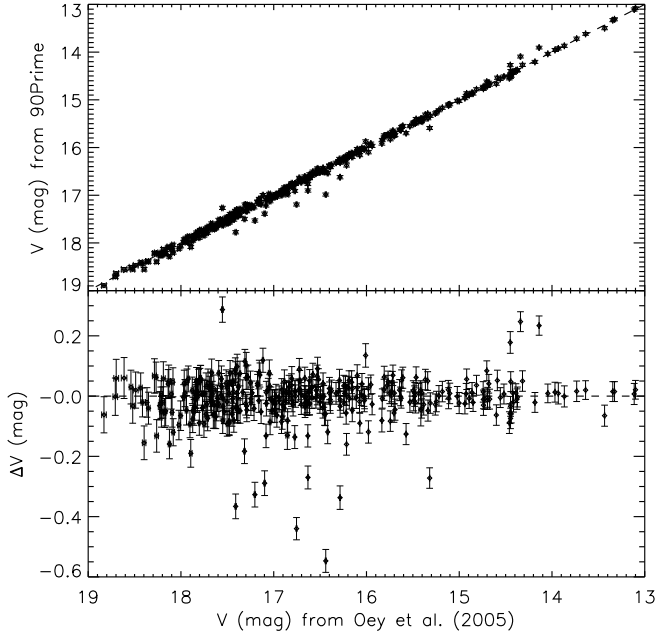}
\caption{Comparison of our 90Prime $V$ magnitudes with those of
  \citet{oey05} for the $354$ stars for which overlapping photometry
  is available. \label{fig:VvsV} }
\end{figure}

We measured $V$ and $I$ photometry for $\sim22,000$ sources, complete
down to $V \sim20$ mag.  In Figure \ref{fig:VvsV}, we compare our $V$
photometry with that of \citet{oey05}, who imaged stars in and around
IC 1795.  Of the $399$ stars in the \citet{oey05} catalog, we have
unsaturated $V$ photometry of $354$, all with $13>V>19$.  Our
magnitudes differ from theirs by a mean of $-0.01$ mag, with an rms
scatter of $0.08$ mag.  We thus consider our photometry to be in good
agreement with that of \citet{oey05}.

\subsection{Spectroscopy}
\label{subsec:spec}

\begin{figure*}
\includegraphics[width = 180mm,trim = 10mm 0 0 0 ]{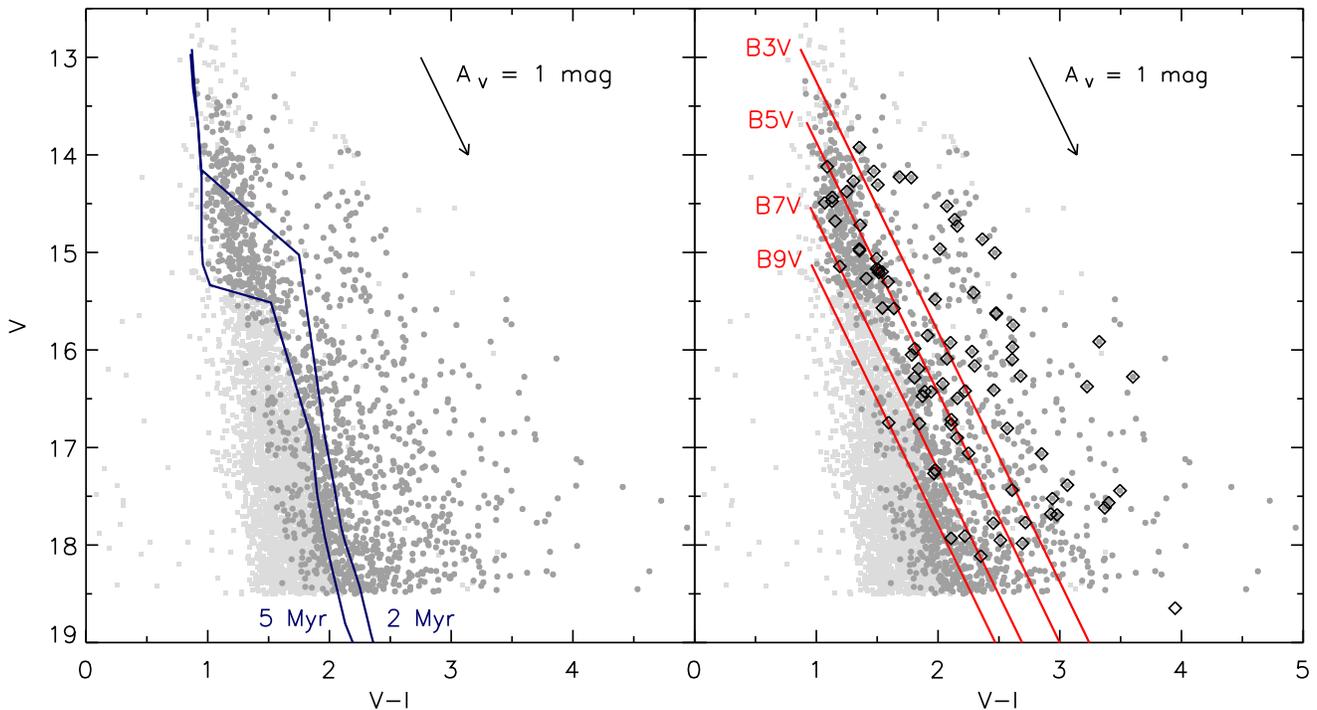}
\caption{Left: $V, V-I$ color-magnitude diagram of 90Prime sources in
  W3 with $V\le18.5$ mag.  Sources with Hectospec spectra are denoted
  by dark grey filled circles, while 90Prime sources that were not
  targeted for spectrosopic follow-up are shown as light grey filled
  squares.  The solid blue lines are the reddened $2$ and $5$ Myr
  pre-main-sequence isochrones from \citet{sie00} that were used in
  target selection.  Right: The same color-magnitude diagram with
  identified O- and B-type stars (see Section \ref{sec:sample})
  highlighted with open diamonds.  Overplotted are the reddening
  tracks (red lines) for B3, B5, B7, and B9 V stars using the
  extinction law from Section \ref{sec:extinct}.
\label{fig:CMD} }
\end{figure*}

We targeted candidate members of W3 for spectroscopic follow-up with
the Hectospec multi-fiber spectrograph \citep{fab05} on the 6.5-m MMT.
Spectroscopic targets were selected on the basis of their location in
the $V$, $V-I$ color-magnitude diagram (Figure \ref{fig:CMD}) relative
to the \citet{sie00} solar-metallicity pre-main-sequence (PMS)
isochrones.  For the purposes of target selection, the PMS isochrones
were corrected for a distance of $2.0$ kpc and a foreground extinction
of $A_V=2.6$ mag.  Sources lying redward of the $5$ Myr PMS isochrone
were prioritized when creating Hectospec fiber pointings, with sources
redward of the $2$ Myr PMS isochrone given the highest priority.
These target selection criteria were designed to minimize the number
of foreground stars observed spectroscopically: a model foreground
population from the Besan\c{c}on Galactic model \citep[][assuming an
  average extinction of $A_V=1.3$ mag kpc$^{-1}$]{rob03}, predicts
that more than $90\%$ ($80\%$) of the expected foreground population
toward W3 lies blueward of the $2$ Myr ($5$ Myr) \citet{sie00}
isochrones.  These target selection criteria also prioritized the
color-magnitude space occupied by O-type and early/mid B-type stars
(see Figure \ref{fig:CMD} and discussion in Section
\ref{subsec:addlit}).

Thirty-one Hectospec fiber configurations were observed in queue mode
over $27$ nights from 2008 to 2012.  Exposure times were chosen to
achieve a minimum signal-to-noise (S/N) ratio of 25 at 6500 \AA.  The
acheived S/N ranged between 22 and 132, with a median of 45.  Details
of each fiber configuation (date of observation, center of field of
view, grating, total on-source exposure time, and number of targets
observed) are given in Table \ref{tab:specdetails}.  A total of 1577
sources were observed one or more times with Hectospec's $270$ lines
mm$^{-1}$ (lpm) grating, which provides $\sim 5$ \AA~resolution at
3650--9200 \AA.  As shown in Figure \ref{fig:CMD}, these sources
included the majority of objects lying redward of the $5$ Myr
\citet{sie00} PMS isochrone with $13<V\le18.5$ mag, as well as several
hundred sources with $V>18.5$ mag (not shown).  A randomly-chosen
subset of 886 targets were also observed with the the $600$ lpm
grating centered on H$\alpha$, which provides $\sim 2$ \AA~resolution
at 5000--7800 \AA.

\capstartfalse
\begin{deluxetable*}{lccccc}
\tablewidth{0pt}
\tablecaption{Hectospec Observing Details \label{tab:specdetails}}
\tablehead{
  \colhead{Date} & 
  \colhead{$\alpha$ (J2000)\tablenotemark{a}} & 
  \colhead {$\delta$ (J2000)\tablenotemark{a}} & 
  \colhead{Grating} & 
  \colhead{Exposure Time} & 
  \colhead{\# Science} 
  \\
  \colhead{(UT)} & 
  \colhead{(HH:MM:SS.s)} & 
  \colhead{($^{\circ}:\arcmin:\arcsec$)} & 
  \colhead{(lines mm$^{-1}$)} & 
  \colhead{(minutes)} & 
  \colhead{Fibers} 
}

\startdata
2008 Dec 4   & 02:26:38.0   & +61:53:36   & 600   & 30    & 176   \\
2008 Dec 4   & 02:26:50.2   & +61:48:11   & 600   & 60    & 183   \\
2008 Dec 5   & 02:26:59.1   & +61:47:15   & 600   & 60    & 154   \\
2008 Dec 7   & 02:26:19.8   & +61:51:29   & 600   & 120   & 217   \\
2009 Feb 3   & 02:27:06.2   & +61:49:57   & 600   & 30    & 179   \\
2009 Feb 3   & 02:27:14.6   & +61:49:57   & 600   & 120   & 213   \\
2009 Feb 5   & 02:26:15.7   & +61:58:46   & 600   & 60    & 101   \\
2009 Feb 5   & 02:27:16.2   & +61:49:13   & 600   & 60    &  99   \\
2009 Mar 2   & 02:27:19.5   & +61:49:42   & 600   & 120   & 213   \\
2009 Oct 11  & 02:26:11.0   & +61:59:19   & 270   & 60    & 205   \\
2009 Oct 16  & 02:26:40.4   & +61:57:01   & 600   & 120   & 215   \\
2009 Oct 17  & 02:26:35.4   & +61:56:55   & 270   & 18    & 207   \\
2009 Dec 12  & 02:27:12.8   & +61:51:10   & 270   & 18    & 178   \\
2009 Dec 15  & 02:27:07.5   & +61:49:41   & 600   & 120   & 163   \\
2010 Oct 2   & 02:26:43.5   & +61:53:22   & 270   & 45    & 160   \\
2010 Oct 3   & 02:28:11.0   & +61:40:37   & 270   & 45    & 152   \\
2010 Oct 10  & 02:27:16.9   & +61:42:32   & 270   & 45    & 183   \\
2010 Oct 10  & 02:26:58.5   & +62:05:53   & 270   & 45    & 128   \\
2010 Oct 11  & 02:25:43.1   & +62:03:14   & 270   & 40    & 139   \\
2010 Oct 16  & 02:25:40.7   & +62:00:25   & 270   & 51    & 128   \\
2010 Oct 17  & 02:25:36.1   & +61:43:10   & 270   & 36    & 134   \\
2010 Nov 26  & 02:27:17.1   & +61:42:08   & 600   & 80    & 186   \\
2010 Nov 27  & 02:25:11.3   & +61:44:38   & 600   & 80    & 224   \\
2011 Sept 20 & 02:25:38.6   & +61:52:04   & 270   & 21    & 154   \\
2011 Oct 24  & 02:25:19.4   & +61:58:54   & 270   & 51    & 172   \\
2011 Oct 25  & 02:26:55.5   & +62:00:35   & 270   & 66    & 107   \\
2011 Nov 17  & 02:26:13.6   & +61:54:21   & 270   & 30    &  84   \\
2011 Nov 18  & 02:25:29.3   & +61:36:18   & 270   & 60    &  54   \\
2012 Jan 23  & 02:27:12.9   & +61:56:09   & 270   & 150   & 222   \\
2012 Feb 11  & 02:25:50.9   & +61:47:28   & 270   & 150   & 231   \\
2012 Feb 12  & 02:26:24.5   & +61:49:21   & 270   & 150  & 222   \\    
\enddata

\tablenotetext{a}{Positions are given for the center of the Hectospec
  field of view for the given fiber configuration.}

\end{deluxetable*}
\capstarttrue

Spectroscopic data were reduced using Juan Cabanela's
E-SPECROAD\footnote{\url{http://astronomy.mnstate.edu/cabanela/research/ESPECROAD}},
an external version of the Hectospec SPECROAD pipeline \citep{min07}.
The E-SPECROAD pipeline automates a set of Hectospec-specific
IRAF\footnote{IRAF is distributed by the National Optical Astronomy
  Observatory, which is operated by the Association of Universities
  for Research in Astronomy (AURA) under cooperative agreement with
  the National Science Foundation.} tasks that perform basic data
reduction: bias and dark-current subtraction, flat-fielding, merging
of four amplifier outputs into one image, aperture extraction, cosmic
ray removal, and combination of multiple exposures.  The pipeline also
corrects for atmospheric water absorption and red light leaks from the
Hectospec fiber-positioning robots.  Wavelength calibration was
performed in IRAF using HeNeAr or HgNeAr calibration lamp
exposures from each night's observations.  As discussed below, we did
not use the E-SPECROAD sky subtraction procedure because the
complexity of the W3 field required a more sophisticated sky
subtraction algorithm.

\subsection{Sky Subtraction \label{subsec:sky}}

\begin{figure*}[t]
\centering
$\begin{array}{cc}
\includegraphics[trim=0mm 0mm 5mm 0mm,clip,scale=0.8]{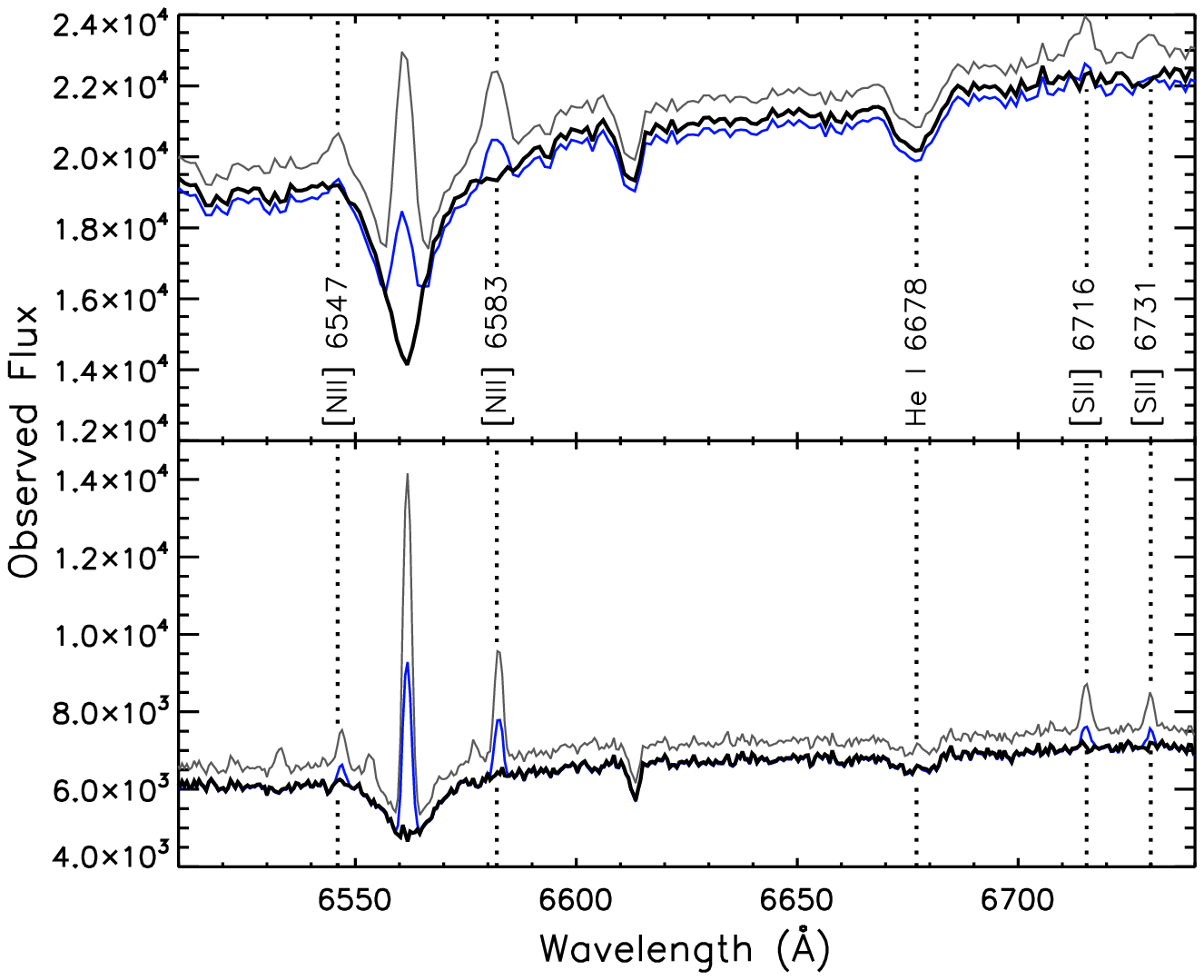} &
\includegraphics[trim=2mm 0mm 0mm 0mm,clip,scale=0.8]{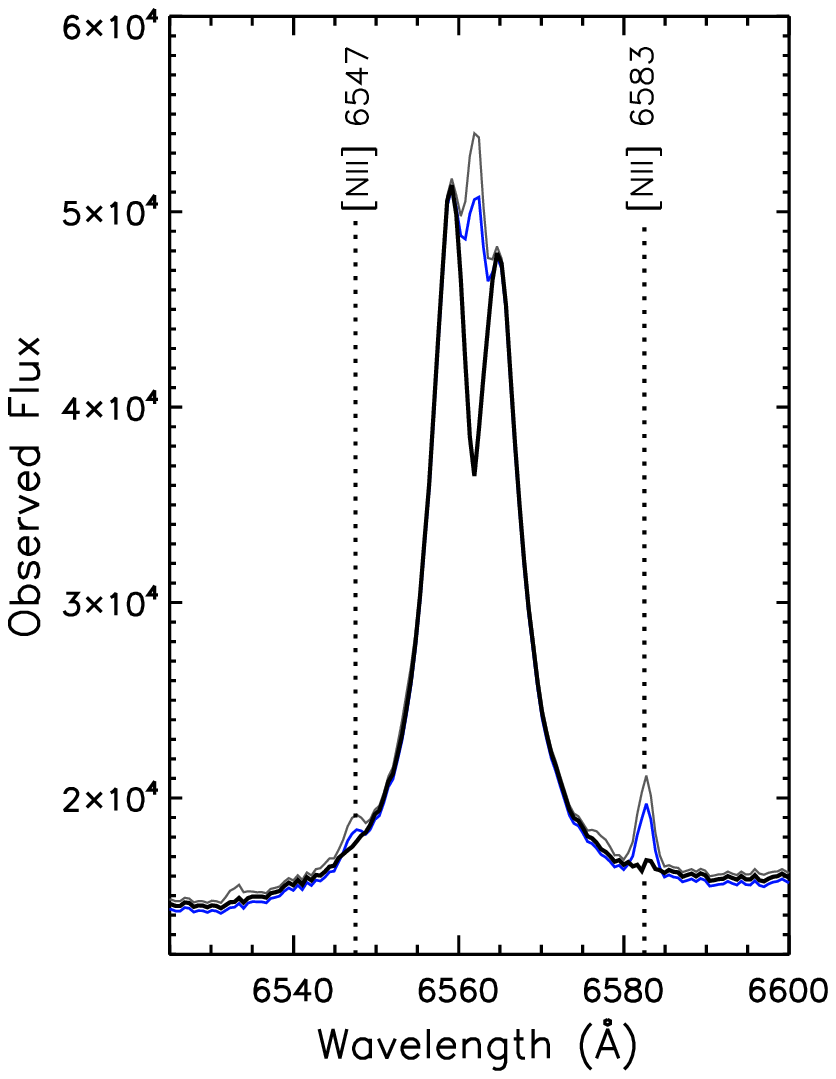} \\
\end{array}$
\caption{Left: Spectra of an example B-type star (source 15 in Table
  \ref{tab:bigtab}), comparing the result of no sky subtraction
  (grey), subtraction of a master sky spectrum (blue), and sky
  subtraction via the method described in Section \ref{subsec:sky}
  (thick black spectrum).  The top spectrum was taken with Hectospec's
  270 lines mm$^{-1}$ (lpm) grating; the bottom spectrum is the same
  object observed with the 600 lpm grating.  Nebular emission lines
  (except for H$\alpha$) and relevant spectral features are labeled.
  Right: Same as left for a source with intrinsic H$\alpha$ emission
  (source 65 in Table \ref{tab:bigtab}).\label{fig:skysub1}}
\end{figure*}

Unlike long-slit spectrographs, fiber spectrographs like Hectospec do
not produce simultaneous observations of the sky emission immediately
adjacent to the science target.  Typically, some fraction of the
fibers in each pointing are placed on blank sky and their spectra are
combined into a master spectrum of sky emission lines, to be
subtracted from every science spectrum.  This method was not suitable
for W3 because of the strong, spatially-variable nebular emission from
its multiple \ion{H}{2} regions.  Each Hectospec fiber subtends
$1.5\arcsec$, capturing the contribution from a small patch of sky
around the target star, including the nebular emission lines. As
illustrated in Figure \ref{fig:skysub1}, incomplete subtraction of
these nebular lines is particularly problematic for the assessment of
the nature of H$\alpha$ absorption/emission in a source.

\begin{figure}
\includegraphics[width = 86mm, trim=9mm 0 2mm 0, clip]{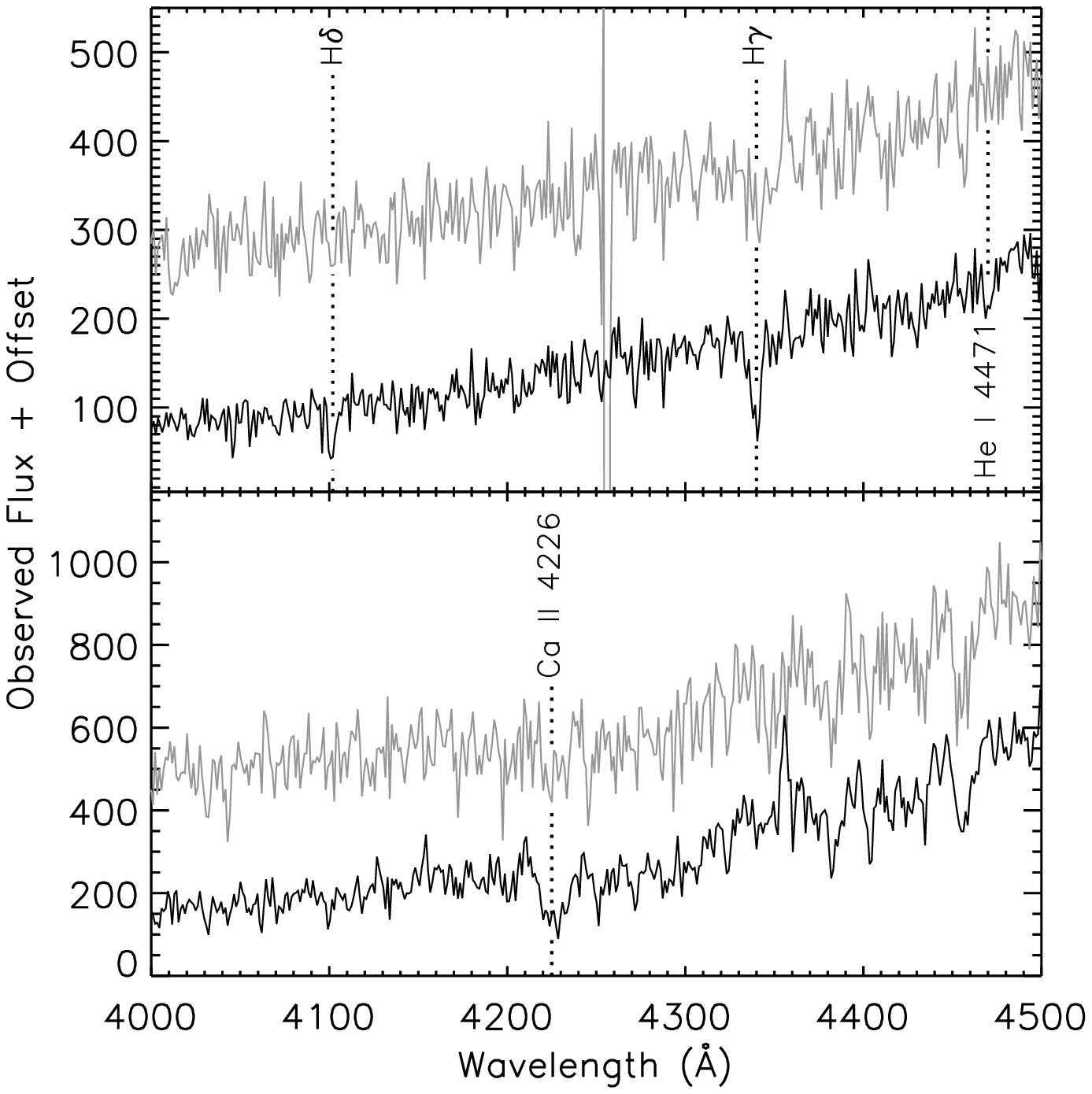}
\caption{Examples of the improvement in signal-to-noise achieved by
  our sky subtraction method (black) compared to directly subtracting
  the sky offset spectra (grey).  The spectra are offset for clarity.
  Some key lines used in the classification of a B-type star (top;
  source 52 in Table \ref{tab:bigtab}) and a K-type star (bottom) are
  marked. \label{fig:skysub2}}
\end{figure}

For each Hectospec pointing, we took an additional exposure shifted
$5$\arcsec~from the science targets, giving each star a corresponding
sky offset spectrum.  To maximize our observing efficiency, the
duration of these sky offset exposures was 1/3 to 1/5 of the total
exposure time spent on-source.  Consequently, as shown in Figure
\ref{fig:skysub2}, directly subtracting the sky offset spectra
produced a noticeable decrease in the S/N of the output.  We therefore
developed our own IDL-based sky subtraction pipeline to maximize the
S/N and accuracy of the resultant spectra.\footnote{The latest version
  of our sky subtraction pipeline is available at
  \url{https://github.com/mkiminki/hectosky}.}

Our sky subtraction pipeline first screens for and excludes any sky
fibers whose spectra are contaminated by nearby stars. All remaining
sky spectra from a given Hectospec pointing are median-combined into a
master sky.  The nebular emission lines (H$\alpha$; H$\beta$;
[\ion{O}{3}] $\lambda\lambda4959,5007$; [\ion{N}{2}] $\lambda\lambda
6547,6583$; [\ion{S}{2}] $\lambda\lambda 6716,6731$; and, when present
above the noise level, \ion{He}{1} $\lambda\lambda 5876,6678,7065$ and
[\ion{Ar}{3}] $\lambda7135$) are fit with Gaussian profiles using the
MPFIT package \citep{mar08} and subtracted from the master sky to
produce a \emph{``master-minus-nebular''} template.  The
\emph{master-minus-nebular} spectrum is subtracted from each
individual sky spectrum, leaving a residual containing only the local
nebular emission lines.  These nebular lines are also fit with
Gaussian profiles, and those profiles are added to the
\emph{master-minus-nebular} template to produce a high-S/N
\emph{``synthetic sky'}' spectrum.  It is this \emph{synthetic sky}
that is subtracted from the corresponding stellar spectrum.

This sky subtraction method accurately subtracts the nebular emission
lines (see Figure \ref{fig:skysub1}) while increasing the continuum
S/N of the output spectra by an average of 30\% relative to directly
subtracting the sky offset spectra (see Figure \ref{fig:skysub2}).
However, as with the simpler options, this method does not accurately
subtract auroral lines such as [\ion{O}{1}] $\lambda\lambda
5577,6300$.  These lines vary too rapidly in time and space to be
accurately removed from spectra taken with a fiber spectrograph.

\begin{figure}
\includegraphics[width = 84mm, trim=2mm 0 10mm 0]{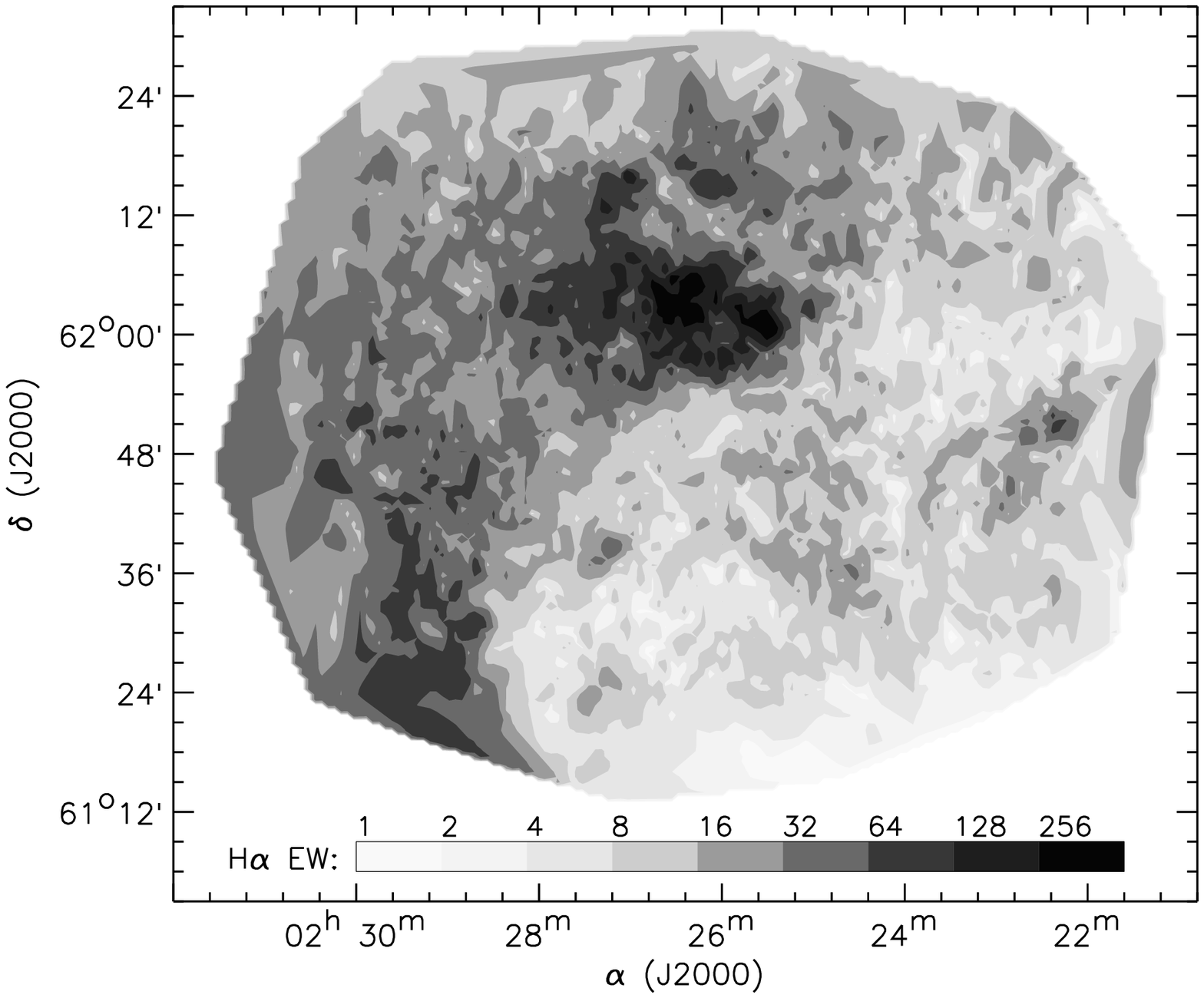}
\caption{Map of H$\alpha$ emission across W3 measured from $6083$
  Hectospec sky spectra ($270$ lpm).  The bright central region is IC
  1795; the left edge of the image is part of the W4 \ion{H}{2}
  region. The equivalent widths have an average uncertainty of
  $\pm1$ \AA.\label{fig:Hamap}}
\end{figure}

An additional product of our sky subtraction pipeline is information
on the level of H$\alpha$ emission across the observed field.  Figure
\ref{fig:Hamap} is a contour map of the nebular H$\alpha$ emission
measured in $6083$ sky spectra.  The \ion{H}{2} region created by IC
1795 is clearly visible near the center of the map, and the
northwest part of the W4 \ion{H}{2} region is evident to the west.

\section{THE O- AND B-TYPE STARS IN W3}
\label{sec:sample}

\subsection{Spectral Classification}

We identified and classified O- and B-type stars via a two-step
process.  First, we screened the spectra of all 1577 Hectospec targets
for the presence of \ion{He}{1} in absorption or emission at $4471$,
$5876$, $6678$, and/or $7065$ \AA.  All stars with plausible
\ion{He}{1} lines were flagged for further analysis.  In cases where
the presence of \ion{He}{1} was debatable, we considered the
progression of the higher-order Paschen lines at 8500--9100 \AA: if
the Paschen lines did not show any blending with the \ion{Ca}{2}
triplet, we flagged the star for further analysis.

This screening procedure resulted in a list of 91 candidate O- and
B-type stars in W3.  It is likely that some of the latest-type B stars
(B8--B9) present in our Hectospec sample were missed, as these stars
exhibit relatively weak \ion{He}{1} absorption.  Very early-type O
stars (O3--O4) also have weak \ion{He}{1} lines and could
theoretically been missed as well; however, given their intrinsic
brightness, it is unlikely that there are any unknown O3--O4 stars at
the distance of W3.  As discussed in Section \ref{subsec:addlit}, the
brightest mid-type O stars in W3 were saturated in our 90Prime
photometry, were not targeted for follow-up spectroscopy, and are
added to our final sample from the literature.

\begin{figure*}
\includegraphics[width = 176mm, trim = 16mm 4mm 8mm 6mm]{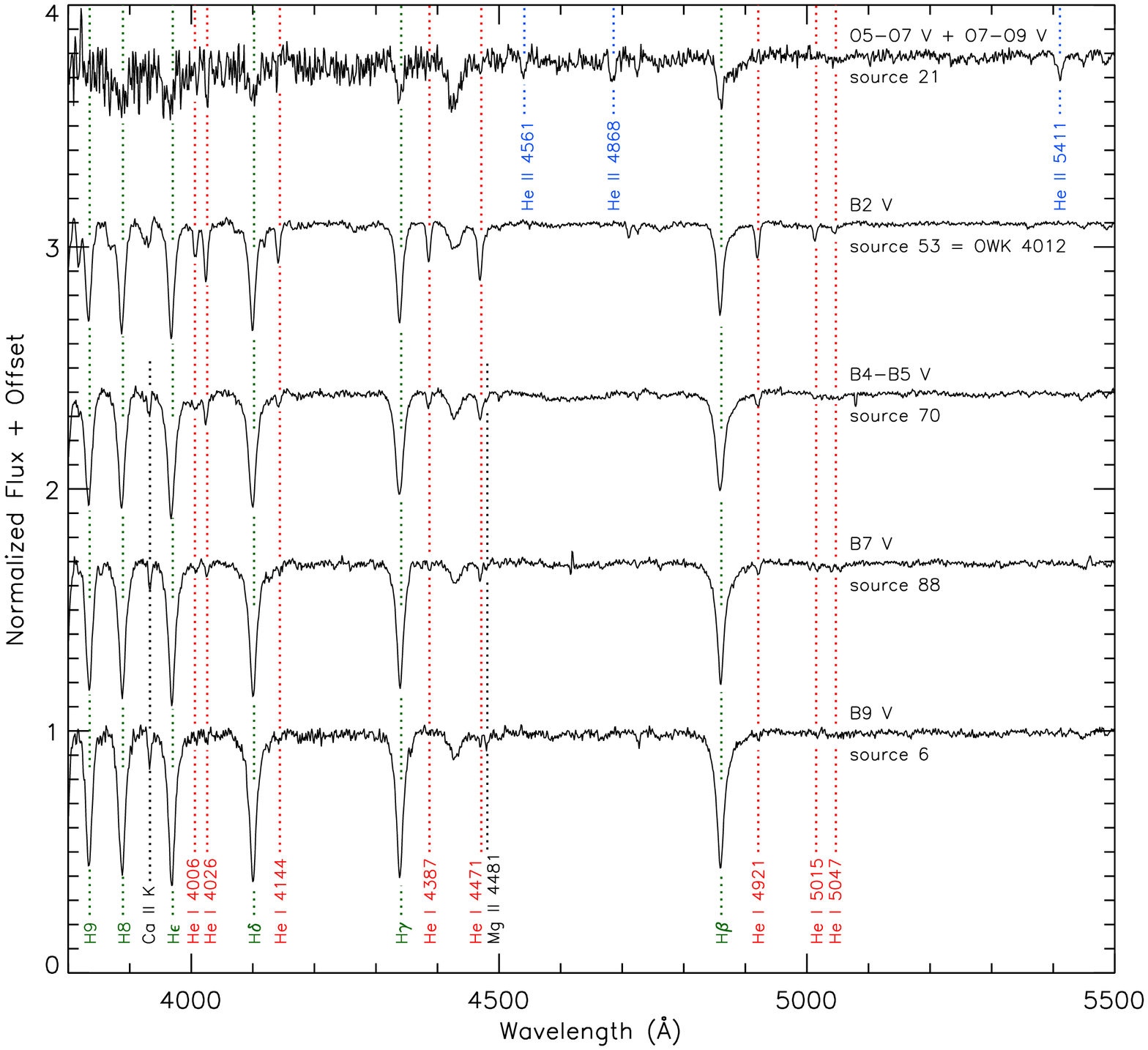}
\caption{Normalized spectra of representative O- and B-type systems in
  W3.  Dotted lines mark the key features used in spectral
  classification (green: hydrogen; red: \ion{He}{1}; blue:
  \ion{He}{2}; black: metals).  All spectra in this figure were taken
  using Hectospec's 270 lines mm$^{-1}$ grating. \label{fig:sptex}}
\end{figure*}

Next, the spectral types of the candidate O- and B-type stars were
determined by visual comparison of their continuum-normalized spectra
to the spectra of standard stars from \citet{gra09}, which are at
comparable resolution ($3.6$ \AA).  Figure \ref{fig:sptex} displays
the 270 lpm spectra of five sources illustrative of the range of
spectral types analyzed.  The key features considered in
classification were: for O-type stars, the ratios of \ion{He}{2} to
\ion{He}{1} line strengths; for all B-type stars, the size and shape
of the \ion{He}{1} lines relative to the hydrogen Balmer lines; and
for B4-B9 stars, the ratio of \ion{He}{1} $\lambda 4471$ to
\ion{Mg}{2} $\lambda 4481$.  The spectral type of most stars in our
sample could be determined to within 1--2 temperature subclasses.  Our
Hectospec spectra did not resolve the metal lines often used for more
precise classification, such as \ion{Si}{4}, \ion{Si}{3}, and
\ion{Si}{2} lines in B0-B3 stars \citep{wal90}.

The spectral resolution was also insufficient to reliably distinguish
between dwarf and giant luminosity classes (we could have identified
supergiants, but found none).  Stars are therefore assumed to be
dwarfs except for one case where multiple authors agreed that the
narrow width of the Balmer lines could only be fit by a giant type.
The effect of this assumption on our derived stellar parameters (see
Section \ref{sec:HRDsect}) is negligible.

\begin{figure*}
\includegraphics[width = 176mm, trim = 10mm 4mm 12mm 6mm]{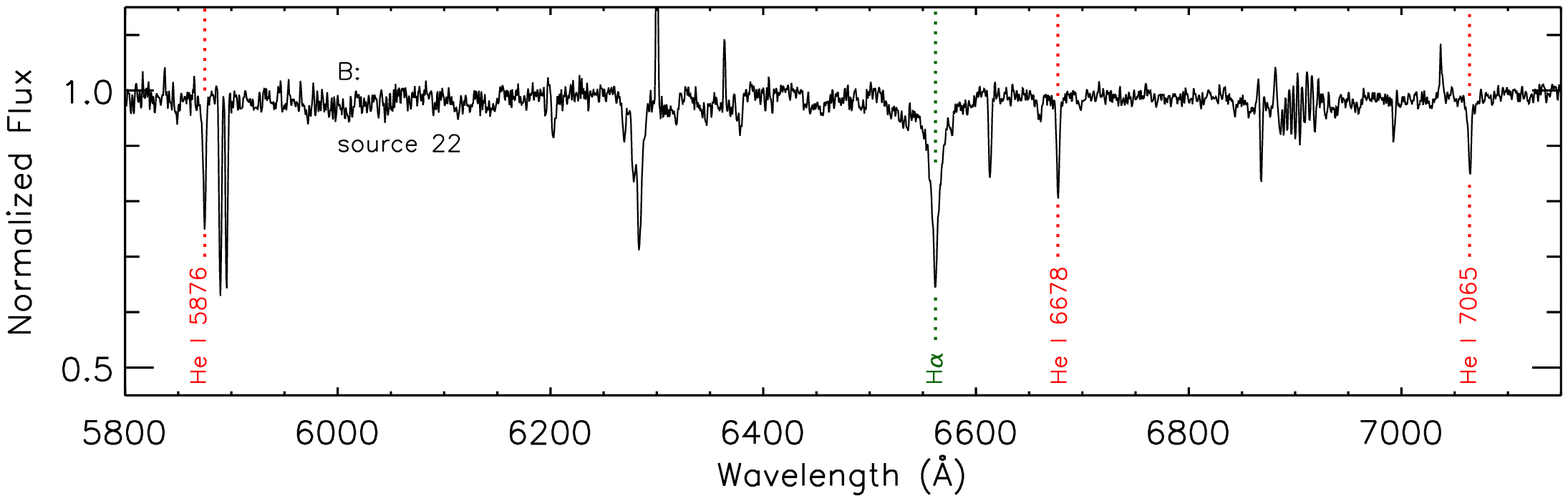}
\caption{Normalized spectrum of a representative B: star, as observed
  with Hectospec's 600 lines mm$^{-1}$ (lpm) grating.  Relevant
  spectral features are marked as in Figure \ref{fig:sptex}.  The B:
  sources cannot be classified using the wavelength range shown in
  Figure \ref{fig:sptex} because their 270 lpm spectra are overly
  noisy or they are heavily extincted in the blue.  Consequently, we
  were unable to determine precise spectral types for these sources,
  although their \ion{He}{1} absorption lines identify them as B-type
  (or possibly late O-type) stars.  Most of the unmarked features in
  this spectrum are of interstellar or atmospheric
  origin.\label{fig:bcol}}

\end{figure*}

Table \ref{tab:bigtab} lists the coordinates, spectral types, and
90Prime photometry of the 91 candidate O- and B-type stars.  Four of
the candidate O- and B-type stars were found to be early A-type stars
and are listed as such.  In addition, we were unable to further
classify ten of the candidate O- and B-type stars because their
spectra were too noisy or too heavily-extincted in the critical
4000--5000 \AA~range.  However, all ten stars exhibited clear
\ion{He}{1} absorption or emission at 6678 and 7065 \AA~(see Figure
\ref{fig:bcol} for an example).  For three of these stars, spectral
types derived from $K_s$-band spectra are available in the literature
(see Section \ref{subsec:addlit}); the remaining seven are listed in
Table \ref{tab:bigtab} as ``B:'' stars.

\subsection{Additional Stars from the Literature}
\label{subsec:addlit}

\citet{oey05}, \citet{nav11}, and \citet{bik12} spectroscopically
identified a combined total of 24 O- and B-type stars in W3.  Our
Hectospec observations included nine of those sources, eight of which
show \ion{He}{1} lines at optical wavelengths and were included in our
initial list of high-mass stars.  We were able to determine spectral
types for five of those eight, and our types are in good agreement
with the literature (within $\pm$ one temperature subtype).  We adopt
the literature classifications for the three in our sample for which
we were unable to determine precise spectral types, and for the $15$
stars not included in our sample.  For spatial completeness, we also
add the two massive stars from \citet{mas95}'s survey of IC 1805 (W4)
that fall within our 90Prime field of view but were not targeted with
Hectospec.

The magnitude distribution of the high-mass stars added from the
literature demonstrates the completeness limits of our Hectospec
sample.  On the bright end, our observations excluded stars with $V<13$
mag, because these sources were saturated in our 90Prime images.  On
the faint side, our completeness varies with spectral type and
extinction (see Figure \ref{fig:CMD}).  Including the O- and B-type
stars detected by \citet{nav11} and \citet{bik12} in $K_s$, which
those groups classified based on their near-infrared spectra, extends
our sample but further complicates an assessment of our completeness,
particularly as those surveys focused on W3 Main and the HDL and did
not explore the rest of the W3 GMC.  We also choose to exclude the
youngest and most embedded objects in W3, as we focus on stars that
have reached or are close to reaching the zero-age main sequence.  For
example, we leave out the three deeply-embedded, massive young stellar
objects (YSOs) identified by \citet{bik12} in W3 Main, along with the
several massive protostars known to be embedded in the infrared-bright
source W3(OH) \citep[e.g.,][]{hir12}.

In total, there are 105 spectroscopically-confirmed O- and B-type
stars in W3.  Of these, 98 have well-constrained spectral types from
this work or the literature, and 92 of those have optical photometry
sufficient to compute their luminosities and place them on a
Hertzsprung-Russell diagram for comparison to stellar evolution models.

\subsection{Infrared Counterparts}
\label{subsec:irmatch}

Using a $1\arcsec$~source-matching radius, we identify the Two Micron
All Sky Survey \citep[2MASS;][]{skr06} $JHK_s$ counterparts to our
sources.  All sources from our data and the literature have 2MASS
counterparts except for IRS N6.  We exclude 2MASS magnitudes with a
signal-to-noise $<5$ (i.e., sources with flags D, E, F, or X) from
further analysis.  For stars in W3 Main, we use the $JHK_s$ photometry
measured by \citet{bik12} with LUCI on the Large Binocular Telescope
for the sources where their photometry is more precise than 2MASS.

In addition, we consider whether any of the O- and B-type stars in W3
show infrared excess at longer wavelengths.  \citet{koe12} used $3.4$,
$4.6$, $12$, and $22~\mu$m photometry from the \emph{Wide-field
  Infrared Survey Explorer} (\emph{WISE}), along with 2MASS $JHK_s$
data, to identify and classify YSOs.  About $2/3$ of the 2MASS
counterparts to the stars in our sample are found in the \citet{koe12}
catalog.  Most of those have \emph{WISE} colors consistent with
stellar photospheres, but eight were classified by \citet{koe12} as
Class I (still enclosed in an infalling envelope) or Class II
(optically thick disk) YSOs (see additional discussion in Section
\ref{subsec:interesting}.)  

\citet{koe12} used a $3\arcsec$ radius to match \emph{WISE} and 2MASS
sources.  The increasingly-large beam size at longer wavelengths means
that source confusion becomes a concern, particularly in crowded
fields like W3 Main.  Two of our sources in W3 Main have apparent
YSO-like infrared excess in \citet{koe12}: \#25 (IRS N2) and \#31 (IRS
N4).  For both sources, the offset between the 2MASS and \emph{WISE}
source positions is $>2.5\arcsec$ and the association is likely
spurious.  Of the remaining six O- and B-type stars across W3 with
YSO-like excesses, two (\#29 and \#60) have positional offsets of
$>1\arcsec$ and so their association with \emph{WISE} sources should
be treated with caution.

While \emph{WISE} has neither the
sensitivity nor the resolution of the \emph{Spitzer Space Telescope},
\emph{WISE} observations cover all of W3, unlike \emph{Spitzer}
surveys, which have mostly been limited to the HDL.  \citet{koe12}
estimate that their sample of Class I and II YSOs in W3 is $ 90 \%$
complete down to $\sim $2--3 M$_\odot$, more than adequate for our
study of the high-mass population.

\subsection{Comments on Spectral Features}
\label{subsec:interesting}

\paragraph{Binaries}

The spectra of stars \#21 and \#61 show the two-component lines
characteristic of double-lined spectroscopic binaries (SB2s).  These
systems have been the target of follow-up observations to characterize
their orbital and physical parameters.  Statistically, most of our
sources are probably binaries (see discussion in Section
\ref{sec:IMFsect}), but our survey was not designed to identify or
characterize source multiplicity.

\paragraph{Be Stars}

Fourteen B stars, including one source from \citet{nav11}, exhibit
definite H$\alpha$ emission, with equivalent widths of 4--55
\AA. (Source \#65, shown in the right panel of Figure
\ref{fig:skysub1}, has a typical level of H$\alpha$ emission for this
group.)  Many of the Be stars display diverse spectral features at
redder wavelengths: five \citep[including the source from][not listed
  in Table \ref{tab:bigtab}]{nav11} have double- or single-peaked
Paschen line emission, three show narrow Paschen line absorption
suggestive of a shell, and one has unusually strong metal absorption
lines. Four Be stars are coincident with \emph{WISE} Class I or II
sources.

\paragraph{Infrared Excesses}

As described in Section \ref{subsec:irmatch}, six B stars, including
four Be stars, are associated with \emph{WISE} Class I or Class II
sources from \cite{koe12}.  The combination of YSO-like infrared
excesses and H$\alpha$ emission in these four Be stars suggests that
they are Herbig Be stars, intermediate-to high-mass PMS stars in which
the H$\alpha$ emission is produced by accretion of material from a
disk onto the star.  However, we cannot rule out the possibility that
these are classical Be stars, rapid rotators with H$\alpha$-emitting
circumstellar disks and near-infrared excesses from free-free
emission.  The two non-H$\alpha$-emitting B stars with \emph{WISE}
infrared excesses are of similarly uncertain evolutionary state, as a
lack of H$\alpha$ emission does not rule out the presence of a disk.

\subsection{Spatial Distribution}

\begin{figure*}
\plotone{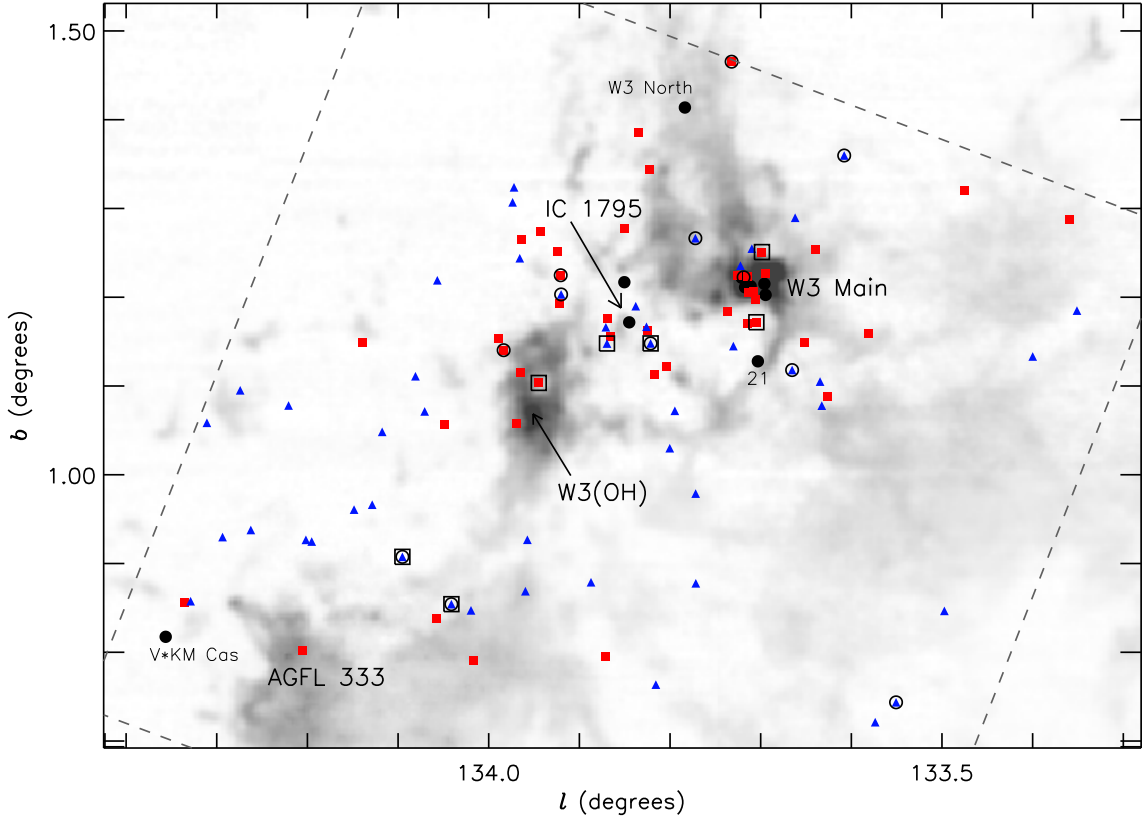}
\caption{Spatial distribution of the O-type (black filled circles),
  early B-type (red filled squares), and mid-to-late B-type (blue
  filled triangles) stars in W3.  The background is $^{12}$CO $J=2$--1
  emission in the velocity range $-60$ to $-25$ km s$^{-1}$
  \citep{bie11}, which highlights the dense gas of the high-density
  layer (HDL) between the W3 GMC and the W4 \ion{H}{2} region.
  Prominent features and specific stars of interest are labelled.  Stars
  with H$\alpha$ emission are marked with open circles, and stars with
  \emph{WISE} IR excess are marked with open squares. The field of
  view of our 90Prime imaging is shown by the dashed
  lines.\label{fig:SPACEfig}}

\end{figure*}

Figure \ref{fig:SPACEfig} compares the distribution of dense gas in the
HDL, as traced by $^{12}$CO $J=2$--1 emission \citep{bie11}, to the
positions of the O- and B-type stars in W3.  While the O-type stars
are generally confined to the HDL, the early B-type (B0--3) and
mid-to-late B-type (B4--9) stars are increasingly present toward the
lower-density interior of the GMC and inside the W4 \ion{H}{2} region.
High-mass stars with \emph{WISE} excess are also concentrated in the
HDL, but several stars with H$\alpha$ emission (also often considered
a marker of youth) are found in the interior of the GMC.  (A
preliminary analysis of the spectra of the lower-mass stars in W3 has
revealed additional H$\alpha$-emitting sources outside the HDL.)  It
is apparent from the observed mass segregation that, while star
formation in W3 is occurring with the greatest vigor in the HDL,
lower-level star formation has been (and may still be) taking place
outside the prominent clusters.

\section{THE EXTINCTION LAW TOWARD W3}
\label{sec:extinct}

Combining our 90Prime photometry with 2MASS data and $B$ photometry
from \citet{oey05} allows us to constrain the wavelength dependence of
extinction for the line-of-sight toward W3.  For each star, we compute
as many as possible of the color excesses $E(B-V)$, $E(V-I)$,
$E(V-J)$, and $E(V-K_s)$.  The intrinsic colors for each spectral type
are taken from \citet{fit70} and \citet{ken95}, with the Johnson $J$
and $K$ bands converted to 2MASS $J$ and $K_s$ using the
transformations of \citet{sie14}.  The instrinsic colors of O-type
stars from the more recent \citet{mar06} study are slightly redder
than those given by other references and do not align with the color
scale for B-type stars.  If we shift our entire intrinsic color scale
to align with that of \citet{mar06}, the color excess ratios remain
the same but the resulting extinctions decrease by an average of
$0.13$ mag.

\begin{figure}[htp]
\includegraphics[width = 84mm, trim = 10mm 0 8mm 0]{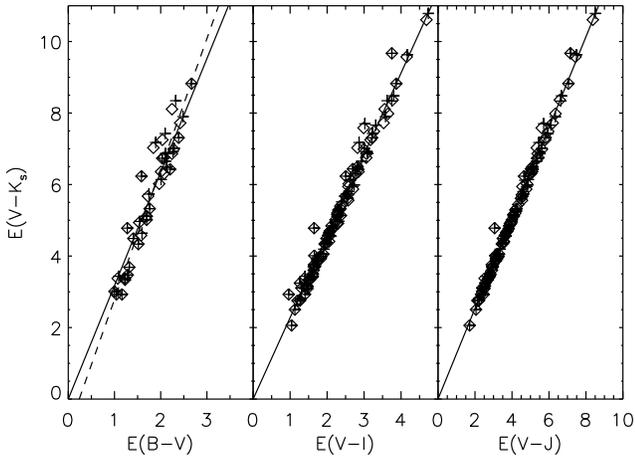}
\caption{$E(V-K_s)$ versus other color excesses for O- and B-type
  stars in W3, used to constrain the extinction law in their
  direction.  Pluses and open diamonds are values assuming a given
  star to be of the earliest and latest spectral type in its range,
  respectively (the two symbols lie atop each other for stars whose
  spectral type could be determined to within one subtype).  Solid lines
  are least-squares fits to the points with the intercept fixed at
  zero, and the dashed line in the left plot is the fit for $E(B-V)$
  versus $E(V-K_s)$ repeated without constraints on the
  intercept. \label{fig:Ecolor}}
\end{figure}

To constrain the extinction at various wavelengths, we consider the
parameter $R_{V\lambda}$, defined here as

\begin{equation}
R_{V\lambda}=\frac{A_V}{E(V-\lambda)} = \frac{1}{1-A_\lambda/A_V}.
\end{equation}

\noindent(We adopt the conventional notation $R_V$ for the value of
$A_V/E(B-V)$.)  It can be shown that

\begin{equation}
R_{V\lambda} = \frac{E(V-K_s)/E(V-\lambda)}{1-(A_{Ks}/A_V)} \ge \frac{E(V-K_s)}{E(V-\lambda)}.
\end{equation}

\noindent In Figure \ref{fig:Ecolor}, we plot $E(V-K_s)$ against the
other color excesses calculated for each star where available.  We
compute the slope of the relationship with a least-squares fit, fixing
the intercept at zero under the assumption that zero extinction in any
one filter means zero extinction in all observed filters.  Only for
the fit of $E(V-K_s)$ against $E(B-V)$ does this assumption have any
effect on the calculated slope.  From the color excess ratios, we find
that $R_V \ge 3.17$, $R_{VI} \ge 2.27$, and $R_{VJ} \ge 1.26$.  We
note that $R_V$ is at least $3.1$, the average for the diffuse
interstellar medium (ISM) in the Milky Way \citep[e.g.,][]{rie85}, and
greater than the $R_V\sim2.9$ found for neighboring W4 by
\citet{han93}. The latter, however, is likely an artificially low
estimate, as they assumed a spectrophotometric distance of 2.4 kpc to
W3 in their analysis.

Lacking independent measures of the absolute magnitudes for any of our
sources, we are unable to put upper limits on $R_{V\lambda}$ from our
data alone.  Instead, we turn to the range of $A_{Ks}/A_V$
measurements in the literature in order to normalize our extinction
law.  Iterating on the $R_V$-dependent \citet{car89} extinction law,
we arrive at $R_V=3.6$ for $A_{Ks}/A_V=0.12$.  Similarly, the
equations of \citet{fit09} produce $R_V=3.61$.  Adopting
$A_{Ks}/A_V=0.12$, we find for W3 that $R_{VI}=2.58$ and
$R_{VJ}=1.43$.  The resulting extinction ratios are
$A_V:A_I:A_J:A_{Ks}=1:0.61:0.30:0.12$.  These ratios are consistent
with the \citet{car89} extinction law for $R_V=3.6$, with the
higher-than-average $R_V$ likely due to the contribution of larger
grains in the W3 GMC.  Using these values, the visual extinction
toward the O- and B-type stars in W3 ranges from $A_V=2.4$ mag to
$12.2$ mag, with an average of $5.7$ mag.

In their study of massive stars in W3 Main, \citet{bik12} found that
the \citet{nis08,nis09} extinction law was the best fit for their
$JHK_s$ data.  This extinction law, derived for sightlines toward the
Galactic Center, has an unusually low $A_{Ks}/A_V$ ratio of $0.062$
\citep{nis08}.  If we adopt that value toward W3, we find that
$R_V=3.38$, $R_{VI}=2.42$, $R_{VJ}=1.34$, and the extinction ratios
are $A_V:A_I:A_J:A_{Ks}=1:0.59:0.25:0.06$.  The resulting $4:1$ ratio
of $A_J:A_{Ks}$ is anomalously steep, even compared to that of
\citet[][who found $A_J:A_{Ks}\sim3:1$]{nis08}.  For the rest of our
analysis, we therefore adopt the more typical $A_{Ks}/A_V=0.12$
described above.  If we do use the lower $A_{Ks}/A_V$ ratio, our
resultant source luminosities decrease by a median of $0.15$ dex, but
our overall conclusions are unaffected.

\section{HERTZSPRUNG-RUSSELL DIAGRAM}
\label{sec:HRDsect}

Using the extinction values from the previous section, we convert our
observed stellar data to physical parameters and construct a
Hertzsprung-Russell (H-R) diagram.  Stellar temperatures are assigned
based on spectral type: temperatures of O-type stars are from the
``observational'' scale of \citet{mar05} and those of B-type stars are
from \citet{ken95} and \citet{sch82}.  Observed $V$ magnitudes are
corrected for extinction and a distance of $2.0$ kpc before bolometric
corrections (BCs) applicable for each spectral type are applied.  We
use BCs from \citet{mar05} for all O-type stars, from \citet{ken95}
for B4-A1V stars, and from \citet{hum84} for BIII and B0-3V stars,
to provide a smooth transition from O-type to B-type dwarfs.

\begin{figure*}
\plotone{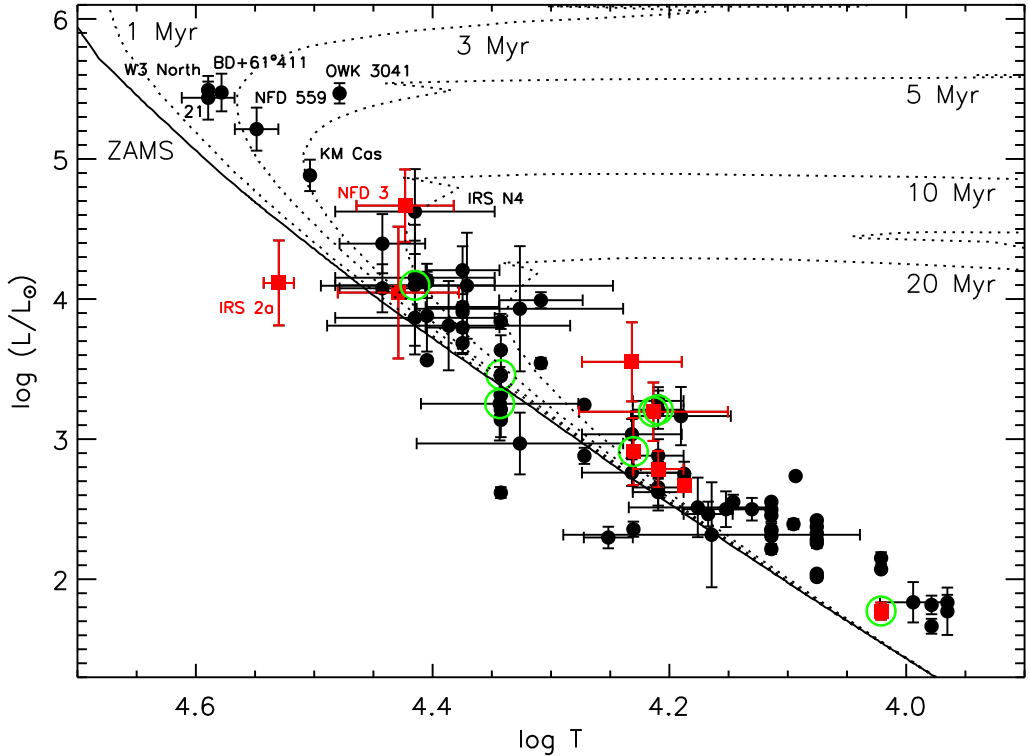}
\caption{H-R diagram of the optically-detected high-mass stars in W3
  for $R_V=3.6$.  Red filled squares and black filled circles are
  stars with and without H$\alpha$ emission, respectively.  Stars that
  show Class I or Class II \emph{WISE} infrared excess according to
  \citet{koe12} are circled in green.  The solid line is the zero-age
  main sequence of \citet{lej01} for a metallicity of $Z=0.02$; the
  dotted lines are the $1$, $3$, $5$, $10$, and $20$ Myr isochrones
  from the same model.
 \label{fig:HRiso}}
\end{figure*}

The final H-R diagram is presented in Figure \ref{fig:HRiso}.
Uncertainites on individual points for both temperature and luminosity
are dominated by the uncertainties in spectral type.  Luminosity
values are also affected by the normalization of the extinction law,
but for clarity we do not plot this systematic uncertainty.  The
majority of the stars are consistent with lying on the main sequence
and there is no obvious main-sequence turnoff.  Overall, there is no
evidence for a population older than $\sim$10 Myr.  Interstingly, the
sources with H$\alpha$ emission and/or \emph{WISE} IR excess do not
occupy any special location in the H-R diagram.

Three of the five hottest optically-detected stars in W3 belong to
known star-forming clusters.  BD+61$^{\circ}$411 is an O6.5V((f)) star
\citep{mat89} that dominates the ionization budget of IC 1795, and NFD
559 \citep{nav11} is the visually-brightest member of W3 Main.  Based
on the positions of BD+61$^{\circ}$411 and OWK 3041 in their H-R
diagram, \citet{oey05} found a 3--5 Myr age spread for IC 1795
assuming $R_V=2.9$ and a spectrophotometric distance of 2.3 kpc.  We
find the same age spread despite using a different distance and
extinction law.  In addition, the $\sim3$ Myr age we estimate for NR
559 is consistent with the 1--3 Myr age for W3 Main found by
\citet{bik12}.  (The youngest members of W3 Main are poorly
represented in our H-R diagram because they are heavily extincted and
therefore undetected in $V$.)

The two hottest optically-detected stars in W3 are W3 North, the
ionizing star of a small \ion{H}{2} region, and our source \#21.
\citet{fei08} did not detect any low-mass stars around W3 North in the
X-ray, leading them to suggest that it was a runaway star ejected from
IC 1795 or IC 1805.  The observed age of $\sim2$ Myr is broadly
consistent with either scenario.  Our source \#21 is one of the
serendipitiously-detected spectroscopic binaries and is composed of
two O-type stars.  This system does not appear to be associated with
any known cluster, although it lies near a dense filament of the HDL
(see Figure \ref{fig:SPACEfig}).  Further study of the surrounding
lower-mass population is necessary before we draw any conclusions
about the origins of source \#21.

The brightest star in W3 with H$\alpha$ emission is NFD 3, a
heavily-extincted object to the southeast of IC 1795.  \citet{nav11}
identified this star as an evolved B0IIIe-B1IIIe based on its $K$-band
spectrum.  Although we observed this star with Hectospec, we were
unable to confirm any spectral type from its optical features.
However, this object's position in the H-R diagram is consistent with
that of a star approaching the end of its main-sequence lifetime.  As
such, it is suggestive of a $\sim10$-Myr-old population in W3.

Another source of note is IRS N4 (our source \#31), an early B-type
star associated with the \ion{H}{2} region W3 K in W3 Main.
\citet{bik12} suggested that IRS N4 is a massive protostar still in
the final stages of contraction to the ZAMS, a hypothesis consistent
with its position on the outskirts of W3 Main but inconsistent with
its visibility in $V$ and its relatively mature \ion{H}{2} region.  It
is possible that IRS N4, like NFD 3, is in fact a somewhat evolved
star that formed before the rest of W3 Main.

A source that stands out in our H-R diagram is IRS 2a, an
H$\alpha$-emitting O-type star lying noticably below the ZAMS.  IRS 2a
is associated with the compact \ion{H}{2} region W3 A in W3 Main
\citep{ojh04} and is therefore unlikely to be a background star.  In
addition, its known near-infrared excess \citep{bik12} may have caused
us to slightly overestimate its extinction: the $A_V$ derived from
$E(V-K_s)$ for IRS 2a is 1.3 mag higher than the $A_V$ derived from
$E(V-I)$.  However, the relatively dense molecular material
surrounding IRS 2a may have a higher $R_V$ than the W3 average.
Placing IRS 2a on or above the ZAMS requires that it have a minimum
$A_V$ of 11.3 mag, which translates to an $R_V$ of 4.0 or greater.  We
note that the source in our sample with the highest measured $A_V$,
star \#27 (NFD 386), is also located in the heart of W3 Main.

Finally, we note that KM Cas, the O-type star responsible for lighting
up a bright-rimmed cloud near AFGL 333, is often considered a member
of IC 1805.  However, with our new determination of its extinction, the
apparent age of KM Cas is $5$ Myr, older than the 1--3 Myr age of the
main IC 1805 cluster \citep{mas95}.  This star is thus also consistent
with the interpretation that star formation began in W3 before the
current OB clusters appeared.

\section{INITIAL MASS FUNCTION}
\label{sec:IMFsect}

\begin{figure}
\includegraphics[width = 84mm, trim = 8mm 0 6mm 0]{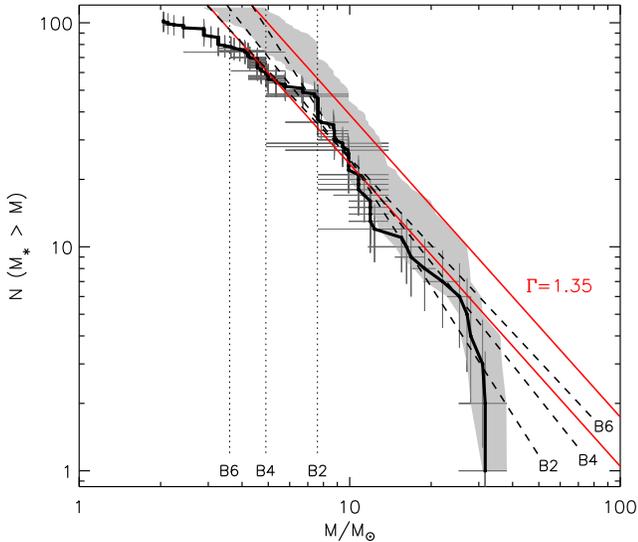}
\caption{Cumulative initial mass function (IMF) for the 102 stars in
  our sample for which masses can be estimated.  Least-squares fits to
  all stars more massive than B2, B4, and B6 dwarfs are shown as
  dashed lines, with the mass cutoffs marked with dotted lines.  The
  grey shaded area illustrates the range of IMFs derived from Monte
  Carlo modelling of the unresolved binary population, assuming a flat
  mass distribution and a binary fraction of 50--90\%.  \citet{sal55}
  IMFs normalized to the observed and binary-corrected IMFs at
  $5$M$_\odot$ are overplotted for comparison (solid red
  lines). \label{fig:IMFfig}}
\end{figure}

We can gain additional insight into the high-mass stellar population
of W3 by exploring the initial mass function (IMF).  We assigned
approximate masses to each star based on its spectral type: O-type
stellar masses are from the ``observational'' scale of \citet{mar05};
masses of unevolved B-type stars are derived by interpolating the
$Z=0.02$ ZAMS from \citet{lej01} to the temperature for each type; and
masses of evolved B-type stars are taken from the mass-spectral type
relations of \citet{dri00}.  None of the stars in W3 are massive
enough to have experienced significant mass loss over their lifetimes
\citep[see, e.g.,][]{smi14}, so the current masses are representative
of the inital masses.  The resulting cumulative mass function is shown
in Figure \ref{fig:IMFfig}.

For stellar masses greater than approximately solar, the IMF is
typically expressed as a power law of the form $dN/d(\log M) \propto
M^{-\Gamma}$, where $\Gamma = 1.35$ is a \citet{sal55} IMF.  We fit
our cumulative histogram in logarithmic space with a straight line of
the form $\log N (M_* > M) = a - \Gamma \log M$ using least-squares
regression and treating the errors on $\log N$ as Poissonian (i.e.,
$\sigma_N = \sqrt{N}$).  The use of a cumulative histogram instead of
a standard one removes the slope dependence on bin size; however, the
fit is still strongly dependent on the mass completeness limit.  If we
assume the sample is complete down to stars of type B6V, we find
$\Gamma=1.29$.  For fits assuming that the sample becomes incomplete for
types later than B4V and B2V, we find $\Gamma=1.52$ and $1.90$,
respectively, as the fit increasingly emphasizes the steep upper end
of the observed IMF.

The observational IMF is almost certainly undercounting the number of
high-mass stars found in unresolved multiple systems.  Half or more --
perhaps even close to $100\%$ -- of massive star systems are multiples
\citep{kim12,san12}.  Furthermore, the mass ratio distribution of
massive systems is roughly uniform, so massive stars are more likely
to have massive companions than if their companions were drawn from a
standard IMF \citep{kim12,san12,kob14}.  Indeed, for two stars (see
Section \ref{subsec:interesting}), we see evidence for similar-mass
companions in the stellar spectra.  Lack of evidence for binarity in
the other observed systems is not evidence for a lack of companions,
as we did not search for radial velocity variations and it was only by
chance that the two SB2 systems were observed near quadrature.

We explore the effect of unresolved binaries on the IMF via a Monte
Carlo simulation.  Each star in our sample is randomly assigned a
binarity status, assuming binary fractions between $50\%$ and $90\%$.
If a binary, we assign a companion mass by drawing from a uniform mass
ratio distribution between $0.005$ and $1.0$.  The range of possible
IMFs produced from $2000$ repetitions of this simulation is shown in
Figure \ref{fig:IMFfig}.  With binaries added, the IMF steepens slightly,
but it remains consistent with Salpeter for $3 \lesssim$ M/M$_\odot$
$\lesssim 30$.  

The drop-off of the observed IMF above $30$ M$_\odot$ is likely the
result of small-number statistics.  The multi-cluster nature of W3 may
also be responsible, since the relationship between cluster mass and
the mass of its most massive star causes the observed high-mass IMF to
steepen in composite populations \citep{wei05}.  However, IC 1795 is
old enough that a star of mass $>40$ M$_{\odot}$, whose presence would
fill out the top end of the W3 IMF, could have already gone supernova.

Assuming that the high-mass IMF in W3 is well-described by a Salpeter
slope, we can extrapolate to estimate the size of the
intermediate-mass stellar population.  Normalizing a Salpeter IMF to
account for unresolved binaries and the seven B-type stars without
firm spectral types, we calculate that there are
$\sim$900--1000 stars with M $>1 $ M$_\odot$ across W3.  In
comparison, \citet{roc11} identified $289$ members and $340$ candidate
members of IC 1795 based on X-ray and IR data.  Excluding the $100$ or
so of their sources with M $<1 $M$_\odot$, these numbers imply that
30--50$\%$ of the stars across our square-degree field of W3 are
concentrated in a single cluster of radius 6\arcmin~in the HDL.

\section{DISCUSSION AND CONCLUSIONS}
\label{sec:disc}

With a wide-field imaging survey and multi-fiber spectroscopy, we have
characterized 109 O- and B-type stars in the W3 star-forming region,
including 83 previously unclassified sources (four of which are A-type
stars). We have shown that the slope of the high-mass IMF in W3 is
consistent with Salpeter ($\Gamma=1.35$), and that the extinction
toward W3 approximately follows the \citet{car89} extinction
law with $R_V\sim3.6$.

Our detection of a population of B-type stars outside the HDL adds to
the growing collection of evidence indicating that star formation has
taken place throughout the W3 GMC.  Our results corroborate the
existence of the $\sim8$-Myr-old population detected in the $K$-band
luminosity function by \citet{rom11}, and we have identified a more
distributed and possibly older population than that seen by
\citet{riv11} and \citet{tow14}.  Diffuse stellar populations of this
nature are common among star-forming regions of all sizes, ranging
from Taurus \citep{sle06} to Orion \citep[e.g.,][]{bri05} to the
Carina Nebula Cluster \citep{fei11}, where more than half of the
X-ray-identified young stars are found spread across a $50$-pc region
outside groups and clusters.

The H-R diagram of the O- and B-type stellar population in W3 also
supports the interpretation that low-level star formation began
throughout the W3 cloud 8--10 Myr ago.  The timing of this initial
star formation may have coincided with the first star formation
episode in neighboring W4, which appears to have occurred 6--10 Myr
ago \citep{den97,oey05}.  However, it is clear from the spatial
distribution of both the observed massive stars and the extrapolated
intermediate-mass population that the star formation in W3 has been
highly concentrated in the HDL over the past 3--5 Myr.  Although the
association of increased star formation rates with dense gas
surrounding an expanding superbubble is not direct evidence that the
latter triggered the former \citep{dal15}, it is very suggestive of a
causal relationship between feedback and accelerated star formation.

It remains possible that the observed distributed population in W3 is
the remnant of an older cluster that has since dispersed.  At the
distance of W3 ($2.0$ kpc), a star with a transverse motion of $2$ km
s$^{-1}$ will cover $\sim17\arcmin$ in $10$ Myr, allowing an unbound
cluster to disperse over $2$ deg$^2$.  The presence of the W3 GMC is
an argument against this hypothesis, however, as the natal cloud of an
older cluster should have largely dispersed after $10$ Myr \citep[see,
  e.g.,][]{lei89}.  \emph{Gaia} will be able to test this possibility
directly, as it is expected to measure the proper motions of $V=18$
mag stars to a precision of $50~\mu$as yr$^{-1}$ \citep{per01,lin08},
which translates to $0.5$ km s$^{-1}$ at the distance of W3.

The next step to understanding the origin and history of the
distributed stellar population in W3 is a study of the intermediate-
and low-mass stars across the region.  At 8--10 Myr
after the commencement of star formation, stars less massive than
$\sim2$ M$_\odot$ will still be contracting toward the main sequence
\citep{sie00}, potentially allowing a more precise age determination.
Both early- and late-type B stars are expected to form surrounded by
close groups of lower-mass stars \citep{tes97}; if such groups prove
to be lacking, the argument for a dispersed cluster remnant is
strengthened.  Further analysis of our 90Prime and Hectospec
observations will allow us to dig deeper into this complex
star-forming region.

\acknowledgments

The authors would like to thank Jessy Jose for her assistance with
preliminary 90Prime photometry; Xavier Koenig for helpful discussions
and for providing us with an updated list of \emph{WISE} source
classifications; Perry Berlind, Mike Calkins, and Nelson Caldwell for
assistance with Hectospec observations; and Juan Cabanela for
technical support for the E-SPECROAD pipeline.  MMK thanks Nathan
Smith for valuable feedback on the manuscript.  JSK would like to
thank Michael R. Meyer for helpful discussions.  Support for this
project was provided by the National Science Foundation through
Astronomy and Astrophysics Research Grant AST-0907980. Observations
reported here were obtained at the MMT Observatory, a joint facility
of the University of Arizona and the Smithsonian Institution. This
publication also makes use of data products from the Two Micron All
Sky Survey, which is a joint project of the University of
Massachusetts and the Infrared Processing and Analysis
Center/California Institute of Technology, funded by the National
Aeronautics and Space Administration and the National Science
Foundation.



\clearpage
\LongTables
\begin{landscape}

\begin{deluxetable}{rcccccccccc}
\tablewidth{0pt}
\tablecaption{Photometry and Spectral Types of Candidate 
  O- and B-Type Stars in W3 \label{tab:bigtab}}
\tabletypesize{\scriptsize}
\tablehead{
  \colhead{ID} &
  \colhead{$\alpha$ (J2000)} &
  \colhead{$\delta$ (J2000)} &
  \colhead{2MASS} &
  \colhead{Alternate}  &
  \colhead{$V$} &
  \colhead{$R$} &
  \colhead{$I$} &
  \colhead{$A_V$\tablenotemark{b}} &
  \colhead{Spectral} &
  \colhead{Comment\tablenotemark{c}}
  \\
  \colhead{} &
  \colhead{} & 
  \colhead{} &
  \colhead{ID} &
  \colhead{IDs\tablenotemark{a}} &
  \colhead{(mag)} &
  \colhead{(mag)} &
  \colhead{(mag)} &
  \colhead{(mag)} &
  \colhead{Type} &
  \colhead{}
}

\startdata

1   & 02:22:39.64  & +62:11:53.0  & J02223962+6211528    & \nodata    & 14.72 $\pm$ 0.03     & \nodata              & 13.36 $\pm$ 0.03     &  3.6 $\pm$ 0.1  & B7 V                 & \nodata    \\
2   & 02:22:50.14  & +61:49:48.1  & J02225013+6149476    & \nodata    & 16.43 $\pm$ 0.03     & 15.47 $\pm$ 0.03     & 14.48 $\pm$ 0.03     &  5.3 $\pm$ 0.1  & B4--B7 V             & \nodata    \\
3   & 02:22:54.02  & +62:07:57.7  & J02225403+6207574    & \nodata    & 15.20 $\pm$ 0.03     & \nodata              & 13.66 $\pm$ 0.03     &  4.3 $\pm$ 0.1  & B4--B5 V             & \nodata    \\
4   & 02:22:57.44  & +61:42:53.8  & J02225743+6142534    & \nodata    & 15.99 $\pm$ 0.03     & 15.08 $\pm$ 0.03     & 14.18 $\pm$ 0.03     &  5.0 $\pm$ 0.0  & B5 V                 & e          \\
5   & 02:23:01.71  & +62:17:29.4  & J02230168+6217292    & \nodata    & 15.92 $\pm$ 0.03     & \nodata              & 13.82 $\pm$ 0.03     &  6.0 $\pm$ 0.0  & B2 V                 & \nodata    \\
6   & 02:23:04.47  & +61:41:09.5  & J02230445+6141093    & \nodata    & 16.05 $\pm$ 0.03     & 15.18 $\pm$ 0.03     & 14.27 $\pm$ 0.03     &  4.7 $\pm$ 0.1  & B9 V                 & \nodata    \\
7   & 02:23:17.74  & +61:35:31.4  & J02231772+6135312    & \nodata    & 16.50 $\pm$ 0.03     & 15.48 $\pm$ 0.03     & 14.34 $\pm$ 0.03     &  5.7 $\pm$ 0.2  & B4--B5 V             & e          \\
8   & 02:24:03.54  & +62:16:53.4  & J02240353+6216531    & \nodata    & 14.17 $\pm$ 0.03     & \nodata              & 12.70 $\pm$ 0.03     &  4.2 $\pm$ 0.0  & B3 V                 & \nodata    \\
9   & 02:24:19.60  & +61:27:40.7  & J02241957+6127403    & \nodata    & 16.61 $\pm$ 0.03     & 15.67 $\pm$ 0.03     & 14.69 $\pm$ 0.03     &  4.7 $\pm$ 0.1  & A1 V                 & \nodata    \\
10  & 02:24:25.89  & +62:05:36.9  & J02242588+6205365    & \nodata    & 15.62 $\pm$ 0.03     & 14.27 $\pm$ 0.03     & 13.14 $\pm$ 0.03     &  6.9 $\pm$ 0.1  & B0--B1 V             & \nodata    \\
11  & 02:24:35.01  & +62:00:41.9  & J02243501+6200417    & \nodata    & 15.97 $\pm$ 0.03     & 14.63 $\pm$ 0.03     & 13.36 $\pm$ 0.03     &  7.3 $\pm$ 0.1  & B1.5 V               & \nodata    \\
12  & 02:24:35.84  & +61:59:58.6  & J02243575+6159581    & \nodata    & 17.77 $\pm$ 0.04     & 16.38 $\pm$ 0.03     & 15.05 $\pm$ 0.03     & \nodata         & B:                   & \nodata    \\
13  & 02:24:41.63  & +62:01:27.5  & J02244164+6201271    & \nodata    & 17.98 $\pm$ 0.03     & 16.63 $\pm$ 0.03     & 15.29 $\pm$ 0.03     & \nodata         & B:                   & \nodata    \\
14  & 02:24:50.85  & +61:32:05.2  & J02245083+6132050    & \nodata    & 15.06 $\pm$ 0.03     & 14.33 $\pm$ 0.03     & 13.57 $\pm$ 0.03     &  4.0 $\pm$ 0.1  & B6 V                 & \nodata    \\
15  & 02:24:57.74  & +62:03:33.8  & J02245771+6203335    & \nodata    & 16.80 $\pm$ 0.03     & 15.48 $\pm$ 0.03     & 14.23 $\pm$ 0.03     &  7.2 $\pm$ 0.0  & B2 V                 & \nodata    \\
16  & 02:24:58.73  & +62:01:32.0  & J02245871+6201317    & \nodata    & 17.06 $\pm$ 0.03     & 15.65 $\pm$ 0.03     & 14.21 $\pm$ 0.03     &  8.2 $\pm$ 0.4  & B3--B5 V             & e+Pa       \\
17  & 02:25:06.11  & +61:45:47.7  & J02250608+6145476    & \nodata    & 15.21 $\pm$ 0.03     & 14.47 $\pm$ 0.03     & 13.69 $\pm$ 0.03     &  4.0 $\pm$ 0.1  & B7.5 V               & \nodata    \\
18  & 02:25:06.54  & +61:38:27.3  & J02250651+6138270    & \nodata    & 15.27 $\pm$ 0.03     & 14.57 $\pm$ 0.03     & 13.86 $\pm$ 0.03     &  3.9 $\pm$ 0.1  & B5--B6 V             & \nodata    \\
19  & 02:25:11.32  & +62:09:43.0  & J02251131+6209428    & \nodata    & 15.85 $\pm$ 0.03     & \nodata              & 13.94 $\pm$ 0.03     &  5.5 $\pm$ 0.0  & B2 V                 & \nodata    \\
20  & 02:25:15.03  & +62:16:19.2  & J02251502+6216191    & \nodata    & 16.01 $\pm$ 0.03     & \nodata              & 13.74 $\pm$ 0.03     & \nodata         & B:                   & e          \\
21  & 02:25:18.59  & +62:01:17.2  & J02251857+6201169    & \nodata    & 16.27 $\pm$ 0.03     & 14.48 $\pm$ 0.03     & 12.67 $\pm$ 0.03     & 10.1 $\pm$ 0.2  & O5--O7 V             & +O7--O9 V  \\
22  & 02:25:24.38  & +61:51:27.9  & J02252435+6151277    & \nodata    & 17.69 $\pm$ 0.04     & 16.13 $\pm$ 0.03     & 14.71 $\pm$ 0.03     & \nodata         & B:                   & \nodata    \\
23  & 02:25:28.29  & +62:11:14.2  & J02252827+6211140    & \nodata    & 16.19 $\pm$ 0.03     & \nodata              & 14.35 $\pm$ 0.03     &  5.0 $\pm$ 0.2  & B8 V                 & \nodata    \\
24  & 02:25:31.68  & +62:03:25.2  & J02253167+6203249    & \nodata    & 14.66 $\pm$ 0.03     & 13.52 $\pm$ 0.03     & 12.53 $\pm$ 0.03     &  6.3 $\pm$ 0.1  & B1 V                 & \nodata    \\
25  & 02:25:32.60  & +62:06:59.8  & J02253258+6206596    & IRS N2     & 18.65 $\pm$ 0.04     & \nodata              & 14.70 $\pm$ 0.03     & 10.8 $\pm$ 0.1  & B1--B2 V             & \nodata    \\
26  & 02:25:34.63  & +62:01:40.3  & J02253461+6201401    & \nodata    & 16.47 $\pm$ 0.03     & 15.51 $\pm$ 0.03     & 14.60 $\pm$ 0.03     &  5.1 $\pm$ 0.0  & B7 V                 & \nodata    \\
27  & 02:25:37.49  & +62:05:24.8  & J02253750+6205244    & NFD 386    & 20.26 $\pm$ 0.09     & 17.94 $\pm$ 0.05     & 15.78 $\pm$ 0.03     & 12.2 $\pm$ 0.1  & B0--B2 V\tablenotemark{d}             & \nodata    \\
28  & 02:25:38.48  & +61:39:01.0  & J02253847+6139007    & \nodata    & 14.23 $\pm$ 0.03     & 13.34 $\pm$ 0.03     & 12.45 $\pm$ 0.03     &  5.1 $\pm$ 0.1  & B2 V                 & \nodata    \\
29  & 02:25:38.81  & +62:08:17.0  & J02253880+6208168    & \nodata    & 16.16 $\pm$ 0.03     & 14.90 $\pm$ 0.03     & 13.86 $\pm$ 0.03     &  6.7 $\pm$ 0.5  & B2 V                 & e+sh+II    \\
30  & 02:25:43.34  & +62:06:15.7  & J02254334+6206154    & IRS 2a     & 19.41 $\pm$ 0.05     & 17.68 $\pm$ 0.04     & 15.92 $\pm$ 0.03     & 10.2 $\pm$ 0.6  & O8--O9 V\tablenotemark{e}             & e          \\
31  & 02:25:44.86  & +62:03:41.5  & J02254485+6203413    & IRS N4     & 15.91 $\pm$ 0.04     & 14.19 $\pm$ 0.03     & 12.59 $\pm$ 0.03     &  9.0 $\pm$ 0.2  & B0--B2 V\tablenotemark{e}             & \nodata    \\
32  & 02:25:44.89  & +62:08:15.8  & J02254488+6208155    & \nodata    & 16.43 $\pm$ 0.03     & 15.49 $\pm$ 0.03     & 14.53 $\pm$ 0.03     &  5.3 $\pm$ 0.2  & B4--B5 V             & \nodata    \\
33  & 02:25:47.21  & +61:53:43.3  & J02254720+6153430    & \nodata    & 16.90 $\pm$ 0.03     & 15.80 $\pm$ 0.03     & 14.74 $\pm$ 0.03     &  6.0 $\pm$ 0.1  & B3--B5 V             & \nodata    \\
34  & 02:25:52.21  & +61:56:12.3  & J02255220+6156120    & \nodata    & 16.28 $\pm$ 0.03     & 15.38 $\pm$ 0.03     & 14.48 $\pm$ 0.03     & \nodata         & B:                   & \nodata    \\
35  & 02:26:01.10  & +61:43:23.9  & J02260110+6143237    & \nodata    & 17.77 $\pm$ 0.03     & 16.56 $\pm$ 0.03     & 15.32 $\pm$ 0.03     &  6.6 $\pm$ 0.1  & B5--B7 V             & \nodata    \\
36  & 02:26:05.87  & +61:58:46.7  & J02260587+6158465    & \nodata    & 14.96 $\pm$ 0.03     & 13.97 $\pm$ 0.03     & 12.95 $\pm$ 0.03     &  5.7 $\pm$ 0.2  & B1.5 V               & \nodata    \\
37  & 02:26:07.29  & +62:02:55.1  & J02260729+6202550    & \nodata    & 15.62 $\pm$ 0.03     & 14.91 $\pm$ 0.03     & 14.20 $\pm$ 0.03     &  3.6 $\pm$ 0.3  & A1 V                 & \nodata    \\
38  & 02:26:10.50  & +61:58:02.0  & J02261050+6158018    & \nodata    & 16.71 $\pm$ 0.03     & 15.68 $\pm$ 0.03     & 14.60 $\pm$ 0.03     &  5.9 $\pm$ 0.1  & B3 V                 & \nodata    \\
39  & 02:26:16.90  & +62:07:35.2  & J02261690+6207350    & \nodata    & 17.62 $\pm$ 0.03     & 16.00 $\pm$ 0.03     & 14.25 $\pm$ 0.03     & \nodata         & B:                   & e+pec      \\
40  & 02:26:17.35  & +61:29:42.3  & J02261734+6129420    & \nodata    & 17.95 $\pm$ 0.04     & 16.72 $\pm$ 0.03     & 15.44 $\pm$ 0.03     &  6.8 $\pm$ 0.2  & B3--B9 V             & \nodata    \\
41  & 02:26:18.80  & +61:59:53.1  & J02261880+6159529    & \nodata    & 15.17 $\pm$ 0.03     & 14.41 $\pm$ 0.03     & 13.67 $\pm$ 0.03     &  4.6 $\pm$ 0.5  & B4 V                 & e+Pa+II    \\
42  & 02:26:22.90  & +62:00:37.3  & J02262290+6200370    & \nodata    & 15.41 $\pm$ 0.03     & 14.25 $\pm$ 0.03     & 13.12 $\pm$ 0.03     &  6.5 $\pm$ 0.1  & B1--B2 V             & \nodata    \\
43  & 02:26:23.39  & +61:28:45.3  & J02262338+6128449    & \nodata    & 14.49 $\pm$ 0.03     & 13.97 $\pm$ 0.03     & 13.42 $\pm$ 0.03     &  2.9 $\pm$ 0.0  & B9 V                 & \nodata    \\
44  & 02:26:24.42  & +62:00:50.4  & J02262442+6200501    & \nodata    & 17.52 $\pm$ 0.04     & 16.03 $\pm$ 0.03     & 14.58 $\pm$ 0.03     &  7.8 $\pm$ 0.2  & B4--B6 V             & \nodata    \\
45  & 02:26:33.49  & +61:41:16.7  & J02263349+6141164    & \nodata    & 16.42 $\pm$ 0.03     & 15.40 $\pm$ 0.03     & 14.20 $\pm$ 0.03     &  6.2 $\pm$ 0.1  & B3--B5 V             & \nodata    \\
46  & 02:26:34.32  & +62:01:53.0  & J02263433+6201527    & \nodata    & 14.38 $\pm$ 0.03     & 13.78 $\pm$ 0.03     & 13.12 $\pm$ 0.03     &  3.8 $\pm$ 0.1  & B4--B5 V             & \nodata    \\
47  & 02:26:34.39  & +62:19:35.3  & J02263439+6219350    & \nodata    & \nodata              & \nodata              & \nodata              & \nodata         & B0--B4 V             & e+sh       \\
48  & 02:26:41.31  & +61:59:24.6  & J02264129+6159242    & OWK 1007   & 14.27 $\pm$ 0.03     & 13.63 $\pm$ 0.03     & 12.96 $\pm$ 0.03     &  3.9 $\pm$ 0.3  & B2 V                 & \nodata    \\
49  & 02:26:41.74  & +61:58:50.6  & J02264173+6158503    & \nodata    & 17.68 $\pm$ 0.04     & 16.23 $\pm$ 0.03     & 14.75 $\pm$ 0.03     &  7.9 $\pm$ 0.1  & B4--B5 V             & I          \\
50  & 02:26:43.00  & +61:44:34.2  & J02264301+6144339    & \nodata    & 15.18 $\pm$ 0.03     & 14.47 $\pm$ 0.03     & 13.67 $\pm$ 0.03     &  4.2 $\pm$ 0.1  & B5 V                 & \nodata    \\
51  & 02:26:45.71  & +61:59:50.7  & J02264570+6159504    & \nodata    & 16.76 $\pm$ 0.03     & 15.72 $\pm$ 0.03     & 14.66 $\pm$ 0.03     &  5.7 $\pm$ 0.1  & B6--B7 V             & \nodata    \\
52  & 02:26:46.42  & +61:35:39.1  & J02264642+6135387    & \nodata    & 17.44 $\pm$ 0.04     & 15.66 $\pm$ 0.03     & 13.95 $\pm$ 0.03     &  9.5 $\pm$ 0.2  & B0--B4 V             & \nodata    \\
53  & 02:26:46.65  & +62:00:26.2  & J02264664+6200259    & OWK 4012   & 13.92 $\pm$ 0.03     & 13.25 $\pm$ 0.03     & 12.57 $\pm$ 0.03     &  4.0 $\pm$ 0.2  & B2 V                 & \nodata    \\
54  & 02:26:55.49  & +62:10:51.9  & J02265547+6210516    & \nodata    & 15.75 $\pm$ 0.03     & 14.45 $\pm$ 0.03     & 13.13 $\pm$ 0.03     &  7.3 $\pm$ 0.3  & B1 V                 & \nodata    \\
55  & 02:26:56.28  & +62:06:31.5  & J02265627+6206313    & \nodata    & 16.27 $\pm$ 0.03     & 14.94 $\pm$ 0.03     & 13.59 $\pm$ 0.03     &  7.6 $\pm$ 0.2  & B0--B3 V             & \nodata    \\
56  & 02:26:57.50  & +61:38:49.3  & J02265752+6138489    & \nodata    & 17.18 $\pm$ 0.03     & 16.33 $\pm$ 0.03     & 15.39 $\pm$ 0.03     &  4.9 $\pm$ 0.2  & A0 V                 & \nodata    \\
57  & 02:26:57.89  & +61:38:46.6  & J02265789+6138462    & \nodata    & 15.30 $\pm$ 0.03     & 14.53 $\pm$ 0.03     & 13.71 $\pm$ 0.03     &  4.3 $\pm$ 0.0  & B8 V                 & \nodata    \\
58  & 02:27:09.17  & +62:12:53.0  & J02270916+6212528    & OWK 7003   & 14.86 $\pm$ 0.03     & 13.70 $\pm$ 0.03     & 12.50 $\pm$ 0.03     &  7.0 $\pm$ 0.2  & B0--B1 V             & \nodata    \\
59  & 02:27:09.36  & +61:38:41.8  & J02270936+6138415    & \nodata    & 16.75 $\pm$ 0.03     & 15.97 $\pm$ 0.03     & 15.15 $\pm$ 0.03     &  4.4 $\pm$ 0.1  & B9 V                 & e+I        \\
60  & 02:27:09.40  & +61:54:44.0  & J02270938+6154437    & \nodata    & 17.44 $\pm$ 0.03     & 16.16 $\pm$ 0.03     & 14.83 $\pm$ 0.03     &  7.3 $\pm$ 0.2  & B1--B3 V             & I          \\
61  & 02:27:12.91  & +61:51:39.2  & J02271290+6151390    & \nodata    & 14.52 $\pm$ 0.03     & 13.49 $\pm$ 0.03     & 12.45 $\pm$ 0.03     &  6.0 $\pm$ 0.1  & B1--B2 V             & +B3--B4 V  \\
62  & 02:27:13.93  & +61:37:25.5  & J02271392+6137251    & \nodata    & 15.63 $\pm$ 0.03     & 14.43 $\pm$ 0.03     & 13.15 $\pm$ 0.03     &  7.2 $\pm$ 0.2  & B1 V                 & \nodata    \\
63  & 02:27:15.13  & +62:00:15.3  & J02271510+6200151    & \nodata    & 14.73 $\pm$ 0.03     & 13.66 $\pm$ 0.03     & 12.57 $\pm$ 0.03     &  6.1 $\pm$ 0.1  & B2 V                 & \nodata    \\
64  & 02:27:16.04  & +62:00:50.8  & J02271602+6200506    & \nodata    & 16.38 $\pm$ 0.03     & 14.77 $\pm$ 0.03     & 13.15 $\pm$ 0.03     & \nodata         & B:                   & e+Pa       \\
65  & 02:27:20.25  & +62:02:02.6  & J02272023+6202023    & \nodata    & 16.10 $\pm$ 0.03     & 14.82 $\pm$ 0.03     & 13.49 $\pm$ 0.03     &  7.6 $\pm$ 0.6  & B0--B1.5 V           & e+sh        \\
66  & 02:27:21.36  & +61:54:57.3  & J02272133+6154570    & NFD 252    & 17.57 $\pm$ 0.04     & 15.83 $\pm$ 0.03     & 14.16 $\pm$ 0.03     &  9.2 $\pm$ 0.6  & B1--B4 V             & \nodata    \\
67  & 02:27:39.55  & +61:56:33.3  & J02273953+6156330    & \nodata    & 17.39 $\pm$ 0.04     & 15.86 $\pm$ 0.03     & 14.32 $\pm$ 0.03     &  8.6 $\pm$ 0.1  & B0--B2 V             & \nodata    \\
68  & 02:27:40.20  & +62:04:20.0  & J02274017+6204197    & \nodata    & 17.06 $\pm$ 0.03     & 15.94 $\pm$ 0.03     & 14.81 $\pm$ 0.03     &  6.4 $\pm$ 0.1  & B1--B4 V             & \nodata    \\
69  & 02:27:44.58  & +61:40:30.3  & J02274456+6140299    & \nodata    & 16.41 $\pm$ 0.03     & 15.17 $\pm$ 0.03     & 13.95 $\pm$ 0.03     &  6.8 $\pm$ 0.2  & B3--B6 V             & e+Pa+II    \\
70  & 02:27:45.34  & +62:02:08.0  & J02274532+6202078    & \nodata    & 14.44 $\pm$ 0.03     & 13.87 $\pm$ 0.03     & 13.30 $\pm$ 0.03     &  3.2 $\pm$ 0.1  & B4--B5 V             & \nodata    \\
71  & 02:27:48.33  & +62:03:19.8  & J02274831+6203196    & \nodata    & 14.31 $\pm$ 0.03     & 13.55 $\pm$ 0.03     & 12.80 $\pm$ 0.03     &  4.5 $\pm$ 0.0  & B1 V                 & \nodata    \\
72  & 02:27:50.29  & +61:49:51.7  & J02275028+6149514    & \nodata    & 18.11 $\pm$ 0.04     & 16.94 $\pm$ 0.03     & 15.76 $\pm$ 0.03     &  6.4 $\pm$ 0.1  & B2 V                 & \nodata    \\
73  & 02:28:00.86  & +62:05:28.9  & J02280084+6205285    & \nodata    & 17.23 $\pm$ 0.03     & 16.23 $\pm$ 0.03     & 15.25 $\pm$ 0.03     &  5.2 $\pm$ 0.1  & B8 V                 & \nodata    \\
74  & 02:28:01.73  & +61:51:39.8  & J02280172+6151395    & \nodata    & 14.45 $\pm$ 0.03     & 13.97 $\pm$ 0.03     & 13.42 $\pm$ 0.03     &  2.4 $\pm$ 0.2  & A0 V                 & \nodata    \\
75  & 02:28:03.19  & +61:50:11.9  & J02280319+6150116    & \nodata    & 16.35 $\pm$ 0.03     & 15.32 $\pm$ 0.03     & 14.31 $\pm$ 0.03     &  5.5 $\pm$ 0.1  & B7 V                 & \nodata    \\
76  & 02:28:03.28  & +62:06:27.7  & J02280324+6206276    & \nodata    & 14.23 $\pm$ 0.03     & 13.37 $\pm$ 0.03     & 12.54 $\pm$ 0.03     &  4.6 $\pm$ 0.1  & B4--B5 V             & \nodata    \\
77  & 02:28:11.11  & +61:43:03.7  & J02281111+6143034    & \nodata    & 15.57 $\pm$ 0.03     & 14.76 $\pm$ 0.03     & 13.94 $\pm$ 0.03     &  4.5 $\pm$ 0.0  & B8 V                 & \nodata    \\
78  & 02:28:15.35  & +61:52:11.5  & J02281533+6152112    & \nodata    & 17.26 $\pm$ 0.03     & 16.29 $\pm$ 0.03     & 15.30 $\pm$ 0.03     &  5.3 $\pm$ 0.0  & B8 V                 & \nodata    \\
79  & 02:28:16.93  & +61:32:11.0  & J02281694+6132102    & \nodata    & 15.00 $\pm$ 0.03     & 13.79 $\pm$ 0.03     & 12.54 $\pm$ 0.03     &  6.9 $\pm$ 0.1  & B2--B3 V             & \nodata    \\
80  & 02:28:19.51  & +61:42:18.9  & J02281951+6142185    & \nodata    & 16.75 $\pm$ 0.03     & 15.83 $\pm$ 0.03     & 14.91 $\pm$ 0.03     &  4.8 $\pm$ 0.2  & B9--A1 V             & \nodata    \\
81  & 02:28:21.11  & +61:47:53.4  & J02282111+6147531    & \nodata    & 17.91 $\pm$ 0.04     & 16.81 $\pm$ 0.03     & 15.69 $\pm$ 0.03     &  5.9 $\pm$ 0.1  & B4 V                 & \nodata    \\
82  & 02:28:24.03  & +61:58:45.5  & J02282402+6158452    & \nodata    & 15.48 $\pm$ 0.03     & 14.53 $\pm$ 0.03     & 13.51 $\pm$ 0.03     &  5.4 $\pm$ 0.0  & B8 III               & \nodata    \\
83  & 02:28:34.86  & +61:39:16.8  & J02283486+6139165    & \nodata    & 17.93 $\pm$ 0.03     & 16.93 $\pm$ 0.03     & 15.82 $\pm$ 0.03     &  5.7 $\pm$ 0.1  & B3--B4 V             & \nodata    \\
84  & 02:28:38.24  & +61:39:14.7  & J02283823+6139144    & \nodata    & 15.57 $\pm$ 0.03     & 14.84 $\pm$ 0.03     & 14.03 $\pm$ 0.03     &  4.2 $\pm$ 0.1  & B8 V                 & \nodata    \\
85  & 02:28:50.20  & +61:53:01.2  & J02285019+6153009    & \nodata    & 16.09 $\pm$ 0.03     & 15.07 $\pm$ 0.03     & 14.02 $\pm$ 0.03     &  5.9 $\pm$ 0.0  & B2 V                 & \nodata    \\
86  & 02:29:08.84  & +61:38:31.6  & J02290883+6138314    & \nodata    & 14.48 $\pm$ 0.03     & 13.94 $\pm$ 0.03     & 13.35 $\pm$ 0.03     &  3.7 $\pm$ 0.2  & B5 V                 & \nodata    \\
87  & 02:29:15.41  & +61:47:16.3  & J02291539+6147160    & \nodata    & 14.97 $\pm$ 0.03     & 14.34 $\pm$ 0.03     & \nodata              &  3.6 $\pm$ 0.1  & B8 V                 & \nodata    \\
88  & 02:29:22.00  & +61:37:23.6  & J02292199+6137234    & \nodata    & 14.12 $\pm$ 0.03     & 13.58 $\pm$ 0.03     & 13.03 $\pm$ 0.03     &  3.1 $\pm$ 0.0  & B7 V                 & \nodata    \\
89  & 02:29:25.08  & +61:32:34.9  & J02292506+6132347    & \nodata    & 15.14 $\pm$ 0.03     & 14.56 $\pm$ 0.03     & 13.95 $\pm$ 0.03     &  3.4 $\pm$ 0.1  & B7 V                 & \nodata    \\
90  & 02:29:43.79  & +61:47:02.6  & J02294376+6147023    & \nodata    & 14.98 $\pm$ 0.03     & 14.31 $\pm$ 0.03     & 13.63 $\pm$ 0.03     &  3.6 $\pm$ 0.1  & B7 V                 & \nodata    \\
91  & 02:29:54.22  & +61:44:11.2  & J02295419+6144110    & \nodata    & 14.68 $\pm$ 0.03     & 14.11 $\pm$ 0.03     & 13.52 $\pm$ 0.03     &  3.2 $\pm$ 0.1  & B7 V                 & \nodata    \\

\enddata

\tablenotetext{a}{OWK = \citet{oey05}; NFD = \citet{nav11}; IRS = named infrared
  source in W3 Main, see \citet{ojh04} and \citet{bik12}.}

\tablenotetext{b}{For $R_V=3.6$ (see Section \ref{sec:extinct}).}

\tablenotetext{c}{e = H$\alpha$ emission; Pa = Paschen emission; sh = shell
  features; pec = peculiar metal line absorption; I(II) = \emph{WISE}
  Class I(II) YSO from \citet{koe12}.}

\tablenotetext{d}{Spectral type from \citet{nav11}.}

\tablenotetext{e}{Spectral type from \citet{bik12}.}

\end{deluxetable}

\clearpage
\end{landscape}


\begin{thebibliography}{}

\bibitem[Bertoldi(1989)]{ber89} Bertoldi, F. 1989, \apj, 346,735
\bibitem[Bieging \& Peters(2011)]{bie11} Bieging, J. H., \& Peter,
  W. L. 2011, \apjs, 196, 18
\bibitem[Bik et al.(2012)]{bik12} Bik, A., Henning, Th., Stolte, A.,
  et al. 2012, \apj, 744, 87
\bibitem[Brice\~no et al.(2005)]{bri05} Brice\~no, C., Calvet, N.,
  Hern\'andez, J., et al. 2005, \aj, 129, 907
\bibitem[Cardelli et al.(1989)]{car89} Cardelli, J. A., Clayton,
  G. C., \& Mathis, J. S. 1989, \apj, 345, 245
\bibitem[Carpenter et al.(2000)]{car00} Carpenter, J. M., Heyer,
  M. H., \& Snell, R. L. 2000, \apjs, 130, 381
\bibitem[Dale et al.(2015)]{dal15} Dale, J. E., Haworth, T. J., \&
  Bressert, E. 2015, \mnras, 450, 1199
\bibitem[Dennison et al.(1997)]{den97} Dennison, B., Topsana, G. A.,
  \& Simonetti, J. H. 1997, \apj, 474, L31
\bibitem[Drilling \& Landolt(2000)]{dri00} Drilling, J. S., \&
  Landolt, A. U. 2000, in Allen's Astrophysical Quantities,
  ed. A. N. Cox (New York, NY: Springer-Verlag), 381
\bibitem[Elmegreen \& Lada(1977)]{elm77} Elmegreen, B. G., \& Lada,
  C. J. 1977, \apj, 214, 725
\bibitem[Elmegreen(1998)]{elm98} Elmegreen, B. 1998, in ASP
  Conf. Ser. 148, Origins, ed. C.~E. Woodward, J.~Shull, \&
  H.~Thronson (San Francisco, CA: ASP), 150
\bibitem[Fabricant et al.(2005)]{fab05} Fabricant, D., Fata, R., Roll,
  J., et al. 2005, \pasp, 117, 1411
\bibitem[Feigelson et al.(2011)]{fei11} Feigelson, E. D., Getman,
  K. V., Townsley, L. K., et al. 2011, \apjs, 194, 9
\bibitem[Feigelson \& Townsley(2008)]{fei08} Feigelson, E. D., \&
  Townsley, L. K.  2008, \apj, 673, 354
\bibitem[FitzGerald(1970)]{fit70} FitzGerald, M. P. 1970, \aap, 4, 234
\bibitem[Fitzpatrick \& Massa(2009)]{fit09} Fitzpatrick, E. L., \&
  Massa, D. 2009, \apj, 699, 1209
\bibitem[Gray \& Corbally(2009)]{gra09} Gray, R. O., \& Corbally,
  C. J. 2009, Stellar Spectral Classification (Princton, NJ: Princeton
  University Press)
\bibitem[Hachisuka et al.(2006)]{hac06} Hachisuka, K., Brunthaler, A.,
  Menten, K. M., et al. 2006, \apj, 645, 337
\bibitem[Hanson \& Clayton(1993)]{han93} Hanson, M. M., \& Clayton,
  G. C. 1993, \aj, 106, 1947
\bibitem[Hirsch et al.(2012)]{hir12} Hirsch, L., Adams, J. D., Herter,
  T. L., et al. 2012, \apj, 757, 113
\bibitem[Humphreys \& McElroy(1984)]{hum84} Humphreys, R. M., \&
  McElroy, D. B. 1984, \apj, 284, 565
\bibitem[Kenyon \& Hartmann(1995)]{ken95} Kenyon, S. J., \& Hartmann,
  L. 1995, \apjs, 101, 117
\bibitem[Kiminki \& Kobulnicky(2012)]{kim12} Kiminki, D. C., \&
  Kobulnicky, H. A.  2012, \apj, 751, 4
\bibitem[Kobulnicky et al.(2014)]{kob14} Kobulnicky, H. A., Kiminki,
  D. C., Lundquist, M. J., et al. 2014, \apjs, 213, 34
\bibitem[Koenig et al.(2012)]{koe12} Koenig, X. P., Leisawitz, D. T.,
  Benford, D. J., et al. 2012, \apj, 744, 130
\bibitem[Lada et al.(1978)]{lad78} Lada, C. J., Elmegreen, B. G.,
  Cong, H.-I., \& Thaddeus, P. 1978, \apj, 226, L39
\bibitem[Leisawitz et al.(1989)]{lei89} Leisawitz, D., Bash, F. N., \&
  Thaddeus, P. 1989, \apjs, 70, 731
\bibitem[Lejeune \& Schaerer(2001)]{lej01} Lejeune, T., \& Schaerer,
  D. 2001, \aap, 366, 538
\bibitem[Lindegren et al.(2008)]{lin08} Lindegren, L., Babusiaux, C.,
  Bailer-Jones, C., et al.  2008, in IAU Symp. 248, A Giant Step: from
  Milli- to Micro-arcsecond Astrometry, ed. W. J. Jin, I. Platais, \&
  M. A. C. Perryman (Cambridge: Cambridge Univ. Press), 217
\bibitem[Markwardt(2008)]{mar08} Markwardt, C. B. 2008, in ASP
  Conf. Ser. 411, Astronomical Data Analysis Software and Systems
  XVIII, ed. D. Bohlender, P. Dowler \& D. Durand (San Francisco, CA:
  ASP), 25
\bibitem[Martins \& Plez(2006)]{mar06} Martins, F., \& Plez, B. 2006,
  \aap, 457, 637
\bibitem[Martins et al.(2005)]{mar05} Martins, F., Schaerer, D., \&
  Hillier, D. J. 2005, \aap, 436, 1049
\bibitem[Massey et al.(1995)]{mas95} Massey, P., Johnson, K. E., \&
  DeGioia-Eastwood, K. 1995, \apj, 454, 151
\bibitem[Mathys(1989)]{mat89} Mathys, G. 1989, \aaps, 81, 237
\bibitem[Megeath et al.(2008)]{meg08} Megeath, S. T., Townsley, L. K.,
  Oey, M. S., \& Tieftrunk, A. R. 2008, in Handbook of Star Forming
  Regions, Vol. 1: The Northern Sky, ed. B. Reipurth (San Francisco,
  CA: ASP), 264
\bibitem[Mink et al.(2007)]{min07} Mink, D. J., Wyatt, W. F.,
  Caldwell, N., et al. 2007, in ASP Conf. Ser. 376, Astronomical Data
  Analysis Software and Systems XVI, ed. R. A. Shaw, F. Hill, \&
  D. J. Bell (San Francisco, CA: ASP), 249
\bibitem[Navarete et al.(2011)]{nav11} Navarete, F., Figueredo, E.,
  Damineli, A., et al. 2011, \aj, 142, 67
\bibitem[Nishiyama et al.(2008)]{nis08} Nishiyama, S., Nagata, T.,
  Tamura, M., et al. 2008, \apj, 680, 1174
\bibitem[Nishiyama et al.(2009)]{nis09} Nishiyama, S., Tamura, M.,
  Hatano, H., et al. 2009, \apj, 696, 1407
\bibitem[Oey et al.(2005)]{oey05} Oey, M. S., Watson, A. M., Kern, K.,
  \& Walth, G. L. 2005, \aj, 129, 393
\bibitem[Ojha et al.(2004)]{ojh04} Ojha, D. K., Tamura, M., Nakajima,
  Y., et al. 2004, \apj, 607, 797
\bibitem[Perryman et al.(2001)]{per01} Perryman, M. A. C., de Boer,
  K. S., Gilmore, G., et al. 2001, \aap, 369, 339
\bibitem[Polychroni et al.(2012)]{pol12} Polychroni, D., Moore,
  T. J. T., \& Allsopp, J. 2012, \mnras, 422, 2992
\bibitem[Rieke \& Lebofsky(1985)]{rie85} Rieke, G. H., \& Lebofsky,
  M. J. 1985, \apj, 288, 618
\bibitem[Rivera-Ingraham et al.(2011)]{riv11} Rivera-Ingraham, A.,
  Martin, P. G., Polychroni, D., \& Moore, T. J. T. 2011, \apj, 743,
  39
\bibitem[Robin et al.(2003)]{rob03} Robin, A. C., Reyl\'e, C.,
  Derri\`ere, S., \& Picaud, S. 2003, \aap, 409, 523
\bibitem[Roccatagliata et al.(2011)]{roc11} Roccatagliata, V.,
  Bouwman, J., Henning, Th., et al. 2011, \apj, 733, 113
\bibitem[Rom\'an-Z\'u\~niga et al.(2011)]{rom11} Rom\'an-Z\'u\~niga,
  C. G., Meg\'ias-V\'asquez, Alves, J. F., \& Lada, E. A. 2011, in
  Stellar Clusters and Associations: A RIA Workshop on Gaia,
  ed. E. J. Alfaro Navarro, A. T. Gallego Calvente, \& M. R. Zapatero
  Osorio (Granada, Spain), 141
\bibitem[Ruch et al.(2007)]{ruc07} Ruch, G. T., Jones, T. J.,
  Woodward, C. E., et al. 2007, \apj, 654, 338
\bibitem[Salpeter(1955)]{sal55} Salpeter, E. E. 1955, \apj, 121, 161
\bibitem[Sana et al.(2012)]{san12} Sana, H., de Mink, S. E., de Koter,
  A., et al. 2012, Science, 337, 444
\bibitem[Sandford et al.(1982)]{san82} Sandford II, M. T., Whitaker,
  R. W., \& Klein, R. I. 1982, \apj, 260, 183
\bibitem[Schmidt-Kaler(1982)]{sch82} Schmidt-Kaler, Th. 1982, in
  Landolt-B\"ornstein, Group VI, Vol. 2., ed. K. Schaifers \&
  H. H. Voigt (Berlin: Springer-Verlag), 454
\bibitem[Sierchio et al.(2014)]{sie14} Sierchio, J. M., Rieke, G. H.,
  Su, K. Y. L., \& G\'asp\'ar, A. 2014, \apj, 785, 33
\bibitem[Siess et al.(2000)]{sie00} Siess L., Dufour E., \& Forestini
  M. 2000, \aap, 358, 593
\bibitem[Slesnick et al.(2006)]{sle06} Slesnick, C. L., Carpenter,
  J. M., Hillenbrand, L. A., \& Mamajek, E. E. 2006, \aj, 132, 2665
\bibitem[Skrutskie et al.(2006)]{skr06} Skrutskie, M. F., Cutri,
  R. M., Stiening, R., et al. 2006, \aj. 131. 1163
\bibitem[Smith(2014)]{smi14} Smith, N. 2014, \araa, 52, 487
\bibitem[Stetson(2000)]{ste00} Stetson, P. B. 2000, \pasp, 112, 925
\bibitem[Testi et al.(1997)]{tes97} Testi, L., Palla, F., Prusti, T.,
  Natta, A., \& Maltagliati, S. 1997, \aap, 320, 159
\bibitem[Townsley et al.(2014)]{tow14} Townsley, L. K., Broos, P. S.,
  Garmire, G. P., et al. 2014, \apjs, 213, 1
\bibitem[Walborn \& Fitzpatrick(1990)]{wal90} Walborn, N. R., \&
  Fitzpatrick, E. L. 1990, \pasp, 102, 379
\bibitem[Weidner \& Kroupa(2005)]{wei05} Weidner, C., \& Kroupa,
  P. 2005, \mnras, 365, 1333
\bibitem[Westerhout(1958)]{wes58} Westerhout, G. 1958, \bain, 14, 215
\bibitem[Whitworth et al.(1994)]{whi94} Whitworth, A. P., Bhattal,
  A. S., Chapman, S. J., Disney, M. J., \& Turner, J. A. 1994, \mnras,
  268, 291
\bibitem[Williams et al.(2004)]{wil04} Williams, G. G., Olszewski, E.,
  Lesser, M. P., \& Burge, J. H. 2004, \procspie, 5492, 787
\bibitem[Xu et al.(2006)]{xu06} Xu, Y., Reid, M. J., Zheng, X. W., \&
  Menten, K. M. 2006, Science, 311, 54

\end{thebibliography}
\end{document}